\documentclass[aps,twocolumn,showpacs,superscriptaddress,floatfix,nofootinbib]{revtex4-1}
\usepackage{graphicx,graphics}
\usepackage{dcolumn}
\usepackage{amsmath,amssymb,amsfonts}
\usepackage{latexsym,verbatim}
\usepackage{bm}
\usepackage{color}
\usepackage[breaklinks=true,colorlinks,citecolor=blue,linkcolor=blue,urlcolor=blue]{hyperref}
\usepackage{tabularx}

\newcommand{\be}{\begin{equation}}
\newcommand{\ee}{\end{equation}}
\newcommand{\bea}{\begin{eqnarray}}
\newcommand{\eea}{\end{eqnarray}}

\begin{document}
\title{Strong enhancement of superconductivity on finitely ramified fractal lattices}
\author{Askar A. Iliasov}
\email{askar.iliasov@physik.uzh.ch}
\affiliation{Institute for Molecules and Materials, Radboud University, Heyendaalseweg 135, 6525AJ Nijmegen, \mbox{The Netherlands}}
\affiliation{Department of Physics, University of Zurich, 8057 Zurich, \mbox{Switzerland}}

\author{Robert Canyellas}
\email{robert.canellasnunez@ru.nl}
\affiliation{Institute for Molecules and Materials, Radboud University, Heyendaalseweg 135, 6525AJ Nijmegen, \mbox{The Netherlands}}

\author{Mikhail I. Katsnelson}
\email{m.katsnelson@science.ru.nl}
\affiliation{Institute for Molecules and Materials, Radboud University, Heyendaalseweg 135, 6525AJ Nijmegen, \mbox{The Netherlands}}
\author{Andrey A. Bagrov}
\email{andrey.bagrov@ru.nl}
\affiliation{Institute for Molecules and Materials, Radboud University, Heyendaalseweg 135, 6525AJ Nijmegen, \mbox{The Netherlands}}

\begin{abstract}
Using the Sierpi\'nski gasket (triangle) and carpet (square) lattices as examples, we theoretically study the properties of fractal superconductors. For that, we focus on the phenomenon of $s$-wave superconductivity in the Hubbard model with attractive on-site potential and employ the Bogoliubov-de Gennes approach and the theory of superfluid stiffness. For the case of the Sierpi\'nski gasket, we demonstrate that fractal geometry of the underlying crystalline lattice can be strongly beneficial for superconductivity, not only leading to a considerable increase of the critical temperature $T_c$ as compared to the regular triangular lattice but also supporting macroscopic phase coherence of the Cooper pairs. In contrast, the Sierpi\'nski carpet geometry does not lead to pronounced effects, and we find no substantial difference as compared with the regular square lattice. We conjecture that the qualitative difference between these cases is caused by different ramification properties of the fractals.
\end{abstract}
\maketitle
Quantum dynamics on artificial fractal lattices is a new research direction that emerged several years ago as a result of recent developments in experimental techniques such as molecular assembly \cite{mol_ass}, supramolecular templating \cite{sup_templ}, scanning tunneling microscopy \cite{STM} (STM), and high energy beam lithography \cite{lithography}, which opened the way to manufacture nanoscale fractal-shaped atomic arrays in a lab on demand. The interest in fractal quantum systems is driven by their unique combination of features that are not typical for conventional solid-state structures. Foremost, due to their fractional Hausdorff dimension, fractals are geometric entities interpolating between regular one-dimensional wires and two-dimensional layers. Given the decisive role of spatial dimension for the physical properties of materials and engineered structures, this can potentially lead to novel physics impossible in integer dimensions. Discrete scale invariance is another attribute of fractals that can strongly affect their spectra of quantum excitations and observable properties \cite{Askar_spectra_1, Askar_spectra_2, Askar_spectra_3}. Finally, fractality can be viewed as a highly peculiar type of defect distribution in a two-dimensional crystal that has a regular geometric structure and is akin neither to clean translationally invariant lattices nor to disorder.

Theoretically, fractal lattices have been explored primarily at the level of classical phase transitions \cite{Gefen1, Gefen2, Gefen3} and single-particle quantum dynamics, where solid knowledge has already been acquired \cite{Kadanoff, Laskin, quantum_walks, vanVeen}. Electronic band structures of certain artificial fractal crystals have been theoretically derived and measured in experiments \cite{fractal_bands}. The non-equilibrium propagation of light through fractal lattices has been studied, and the mathematical theory of diffusion on them has been developed \cite{fractal_light}. Topologically protected phases were shown both theoretically \cite{fractal_HOT, Neupert_topology, fractal_topology} and experimentally \cite{Canyellas} to exist in fractal structures, and even the existence of a novel class of topological states impossible in regular crystalline lattices -- topological random fractals -- has been conjectured \cite{topological_random_fractals}. Researchers have analyzed electron transport in fractals and related conductance fluctuations in fractal structures to their Hausdorff dimensions \cite{fractal_conductance}, and predicted plasmon confinement in finitely ramified lattices \cite{westerhout_plasmons}.

\begin{figure*}[t!]
    \centering
    \includegraphics[width=0.95\textwidth]{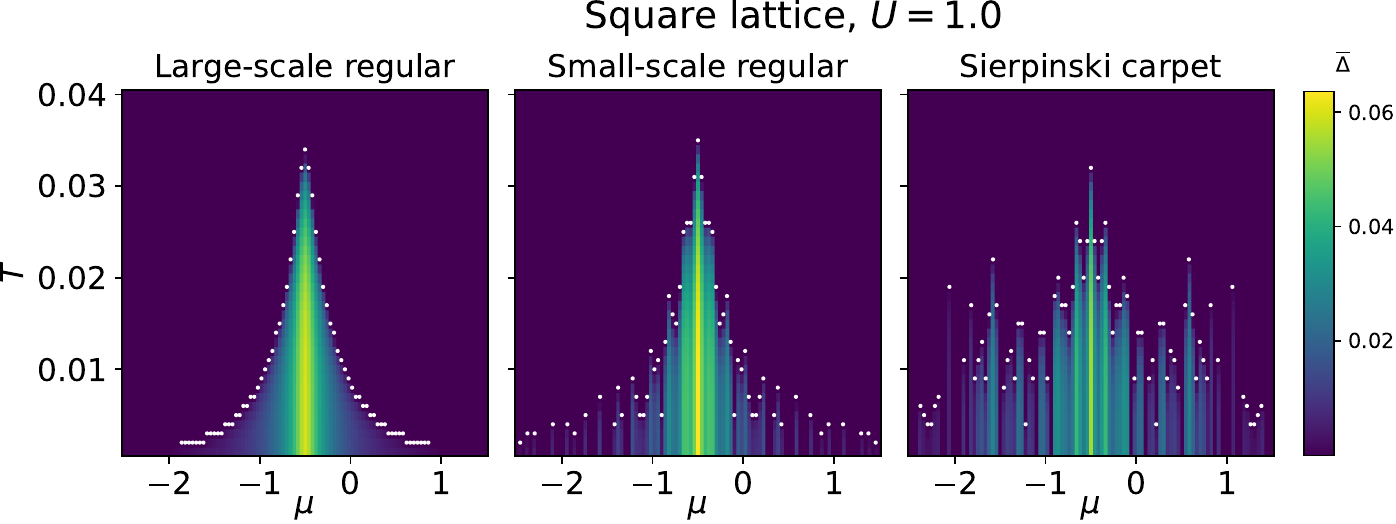}
    \vskip 15pt
    \includegraphics[width=0.95\textwidth]{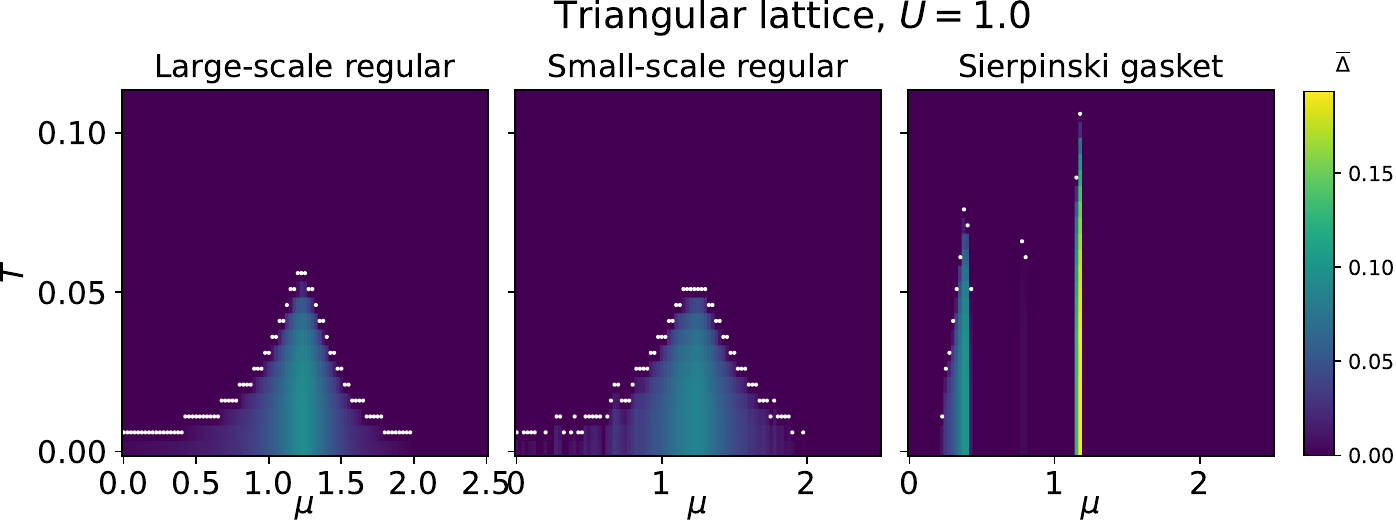}
    \caption{\label{fig:phase_diagram}$T-\mu$ phase diagrams of the attractive Hubbard model on regular lattices and fractal structures. For both the triangular and the square geometries, the BdG equations have been solved self-consistently within the Hartree-Fock approximation for three cases. White dots mark the boundaries of superconducting domes. Left column: large-scale lattices mimicking the thermodynamic limit -- square-shaped samples (for both geometries) with side length $L=1000$ and periodic boundary conditions. Middle column: small-scale samples with regular lattice geometry; for the square lattice, the sample is of square shape (side length $L=27$), and for the triangular one -- of the triangular shape (side length $L=33$), both with open boundary conditions. Right column: Sierpi\'nski carpet (side length $L=27$) and gasket (side length $L=33$) with open boundary conditions.}
\end{figure*}

In the domain of correlated systems, fractals have been explored to much lesser extent. Some results have been obtained regarding {\emph{formation}} of fractal patterns in otherwise regular settings. Fractal Fermi surfaces were shown to emerge in the band structure of non-Hermitian models \cite{nodal_bands}. For driven one-dimensional systems, formation of quantum states with fractional \cite{Khemani} and explicitly fractal-shaped scaling of entanglement entropy \cite{Ageev} was demonstrated. The few undertaken attempts to study fractal geometries {\emph{hosting}} many-body quantum dynamics led to highly promising results, but only particular cases have been investigated. For example, it was shown that the interplay of fractality and interactions can amplify the robustness of the fractional quantum Hall phase \cite{Nielsen-1, Nielsen-2} or host spin liquid phases \cite{fractal_SL}. Auxiliary-field quantum Monte Carlo study of the Sierpi\'nski gasket Hubbard model showed the existence of localized states and ferrimagnetic order \cite{AFQMC}. In the realm of superconductivity, it was shown both theoretically and experimentally that disorder-induced multifractality enhances superconductivity in thin films \cite{Burmistrov-1, Burmistrov-2, multifrac_exp}. However, these results account only for the average scale invariance of disorder, and understanding the possible role of discrete self-similarity and highly symmetric geometry of regular fractal lattices, such as the Sierpi\'nski gaskets and carpets, remains an open problem. Here, we make the first step in this direction and study superconductors with regular fractal geometry. We pursue this analysis motivated by the hope that due to the regular, though non-crystalline, geometric structure of fractals one can hopefully gain benefits of disorder, such as the increase in the critical temperature of formation of individual Cooper pairs \cite{disorder_SC}, without suffering from the destructive self-interference and losing the long-range phase coherence of the superconducting condensate.

For that, we consider the minimal model known to host superconductivity -- the Hubbard model with on-site attractive interactions:
\begin{equation}\label{eq:Hubbard_model}
    H=t\sum_{\langle i,j \rangle\sigma} c^{\dagger}_{i\sigma}c_{j\sigma}-\mu\sum_{i\sigma}n_{i\sigma}-U\sum_{i} n_{i\uparrow} n_{i\downarrow}
\end{equation}
where $c^\dagger_{i\sigma},\,c_{i\sigma}$ are creation and annihilation operators of fermion with spin $\sigma=\uparrow,\downarrow$ at site $i$, $n_{i\sigma}=c^{\dagger}_{i\sigma} c_{i\sigma}$ is the particle number operator, $t=-1$ is the nearest-neighbor hopping, $\mu$ is the chemical potential, and $U>0$ is the strength of attractive on-site interactions.

\begin{figure*}[t!]
    \centering
    \includegraphics[width=0.95\textwidth]{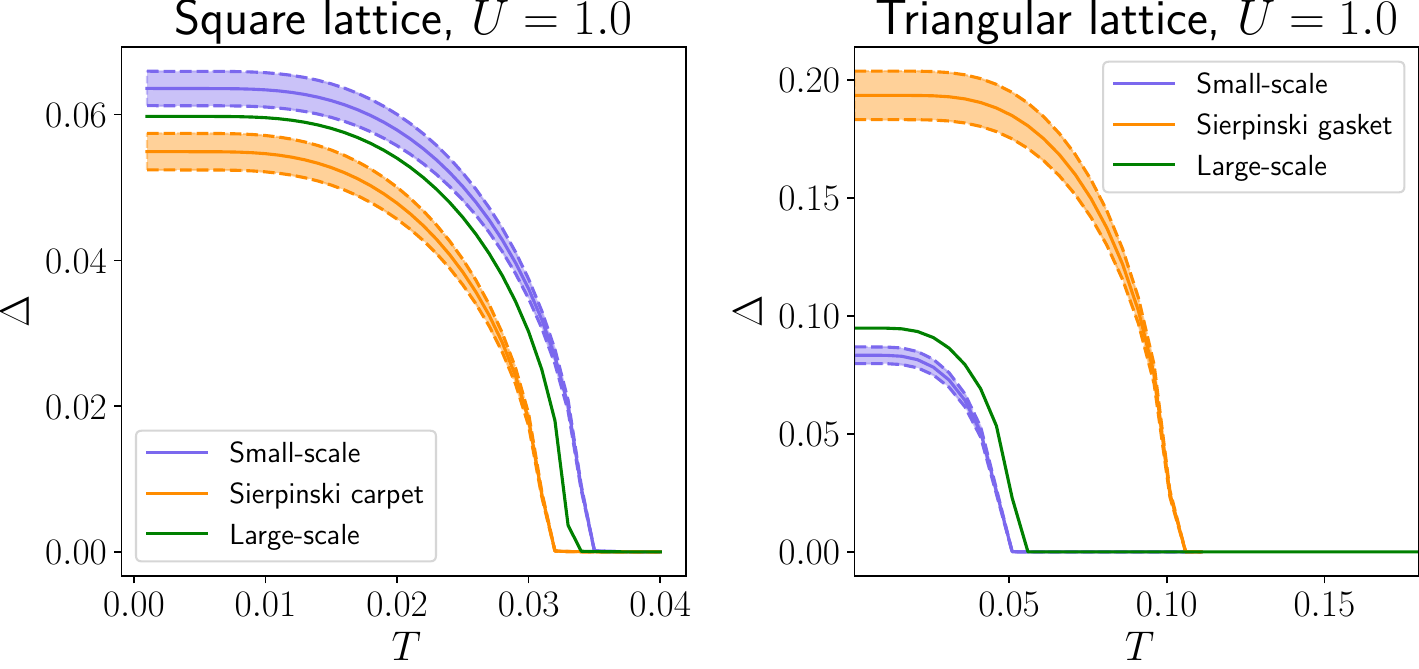}
    \caption{\label{fig:1d_slice}Spatial average of the Cooper pair condensate $\overline{\Delta}$ as a function of temperature $T$ at $U=1$ and optimal chemical potential ($\mu\simeq -0.5$ for the square lattice, and $\mu\simeq 1.2$ for the triangular one). For each parental geometry, three cases are shown -- regular lattice in the thermodynamic limit (green lines; simulated as square-shaped unit cells with side length $L=1000$ and periodic boundary conditions), small-scale sample with regular structure and open boundary conditions (purple lines; simulated as a square cell with side length $L=27$ and right triangular cell with side length $L=33$ correspondingly), and finite-size fractals with open boundary conditions (orange lines; the same side lengths as for regular small-scale lattices). Shaded regions show spatial standard deviations of $\Delta$ from $\overline{\Delta}$ across the lattice.}
\end{figure*}

\begin{figure*}[t!]
    \centering
    \includegraphics[width=0.45\textwidth]{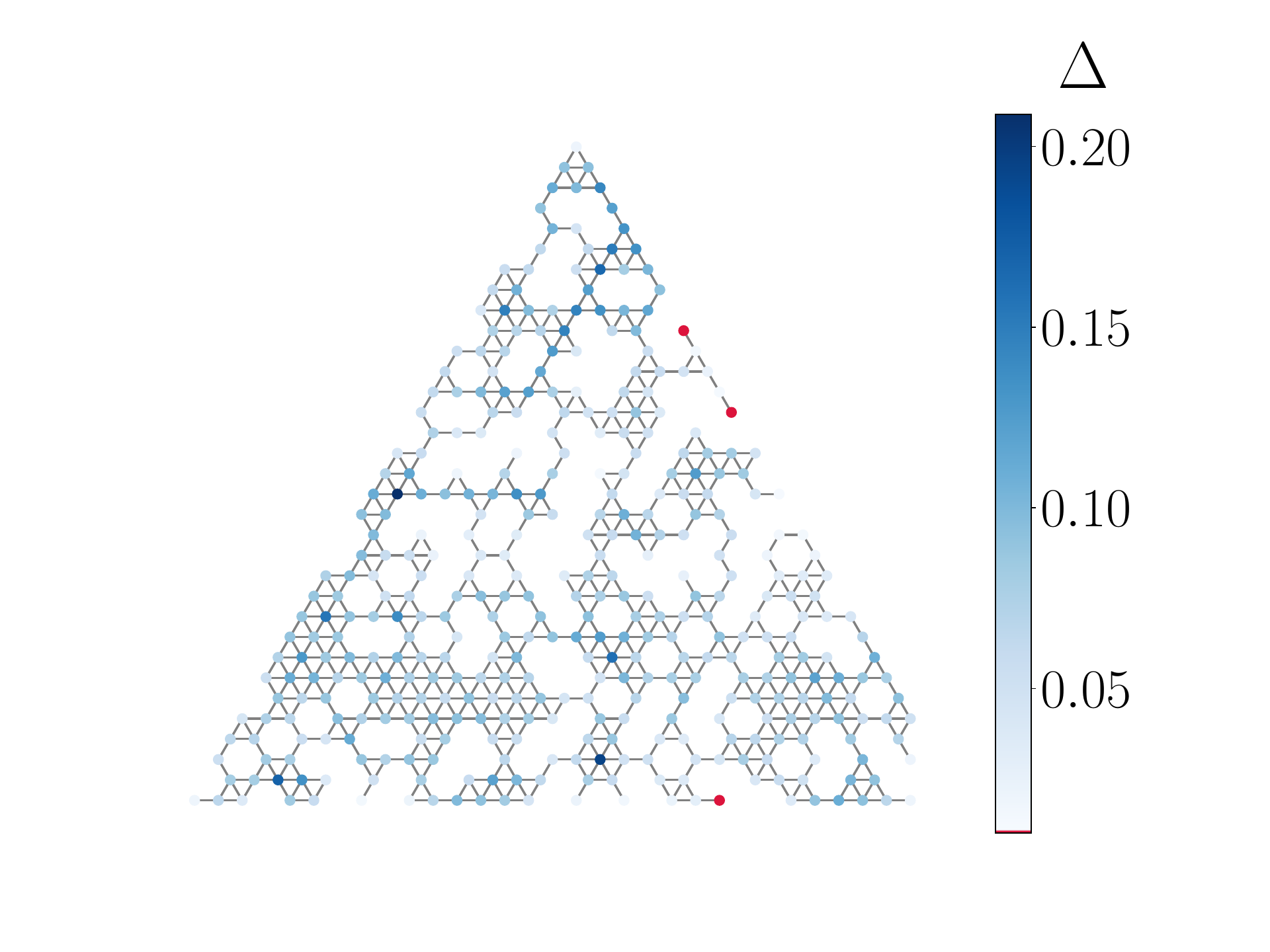}
    \includegraphics[width=0.45\textwidth]{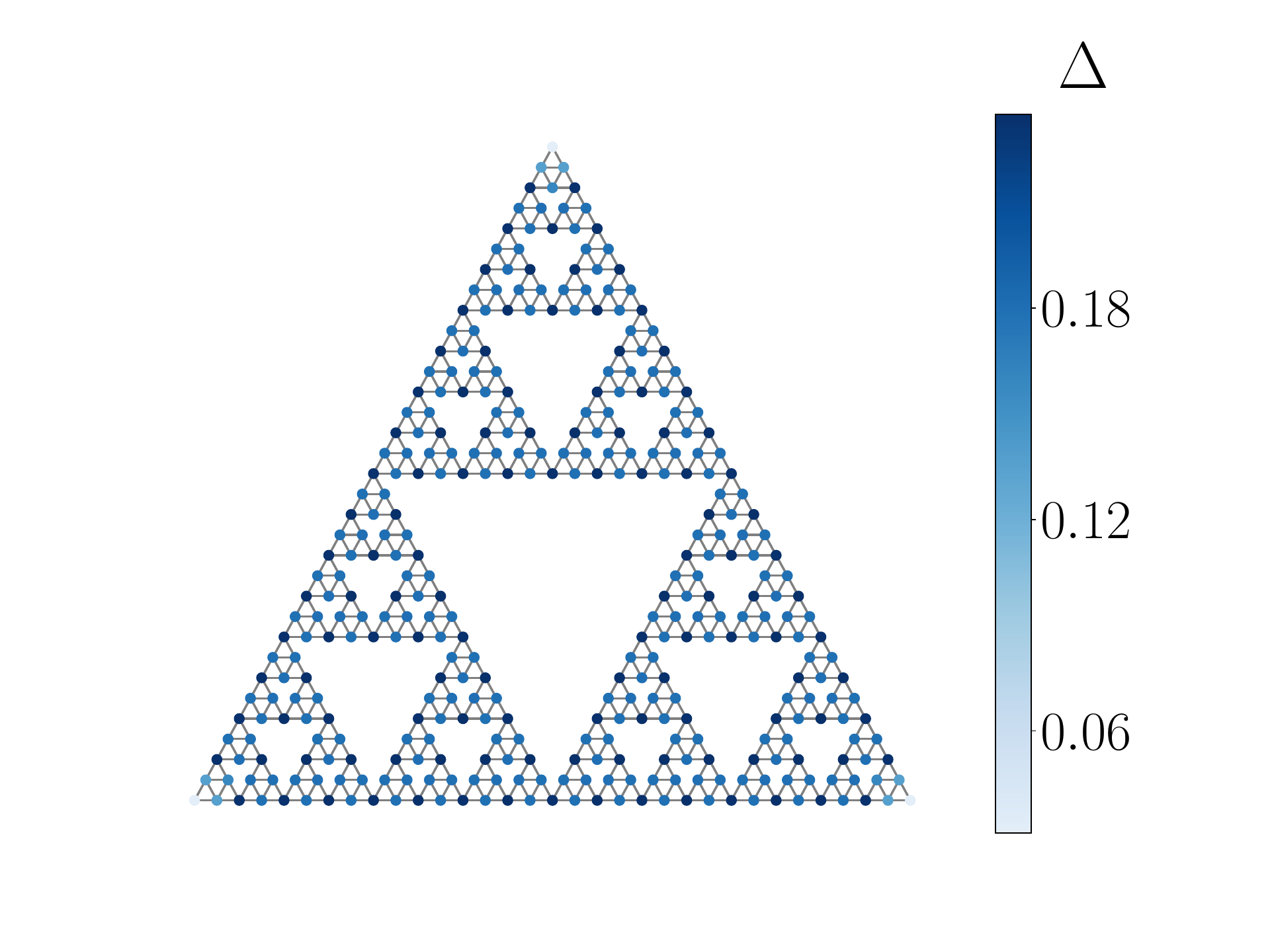}
    \includegraphics[width=0.45\textwidth]{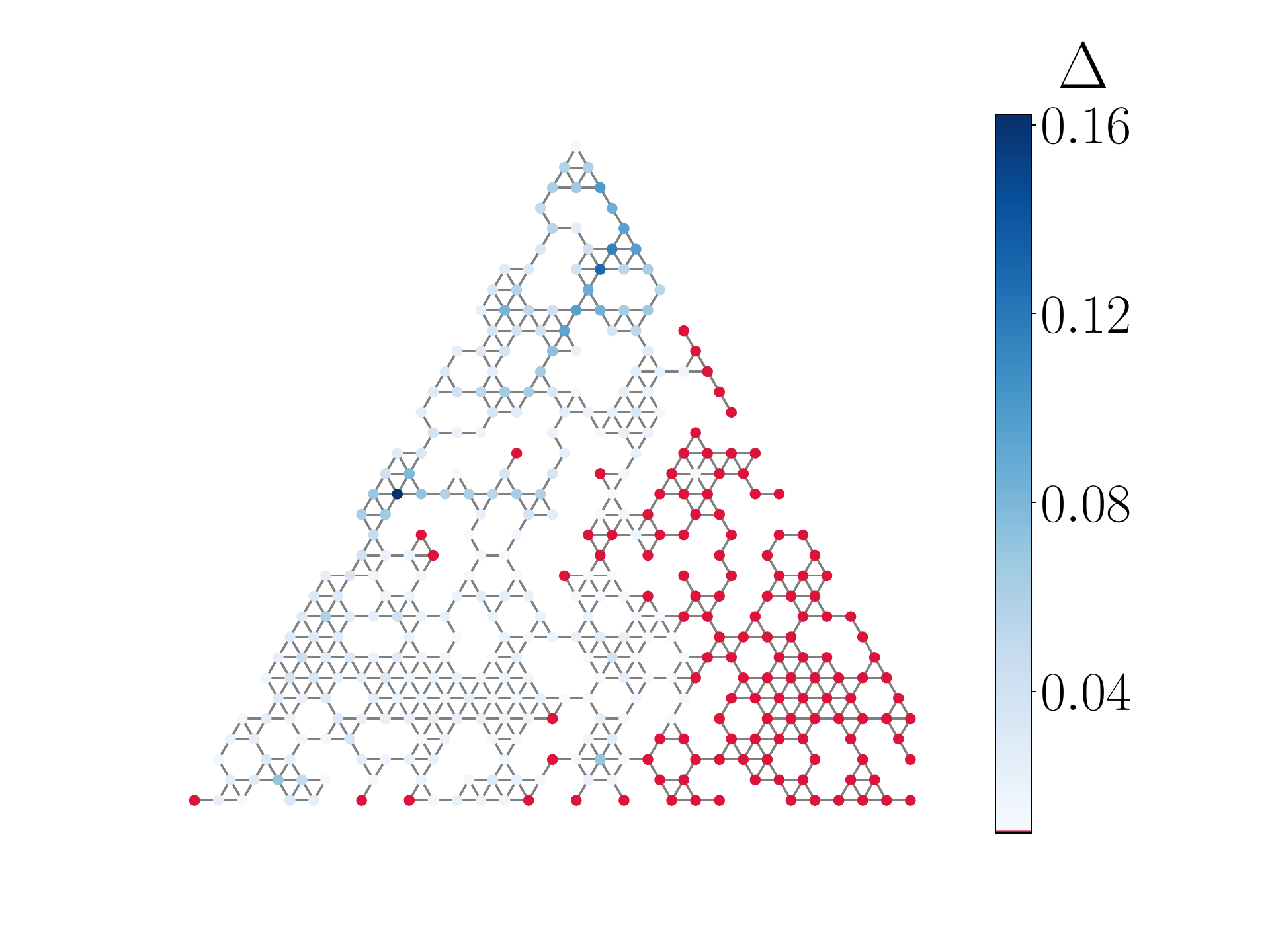}
    \includegraphics[width=0.45\textwidth]{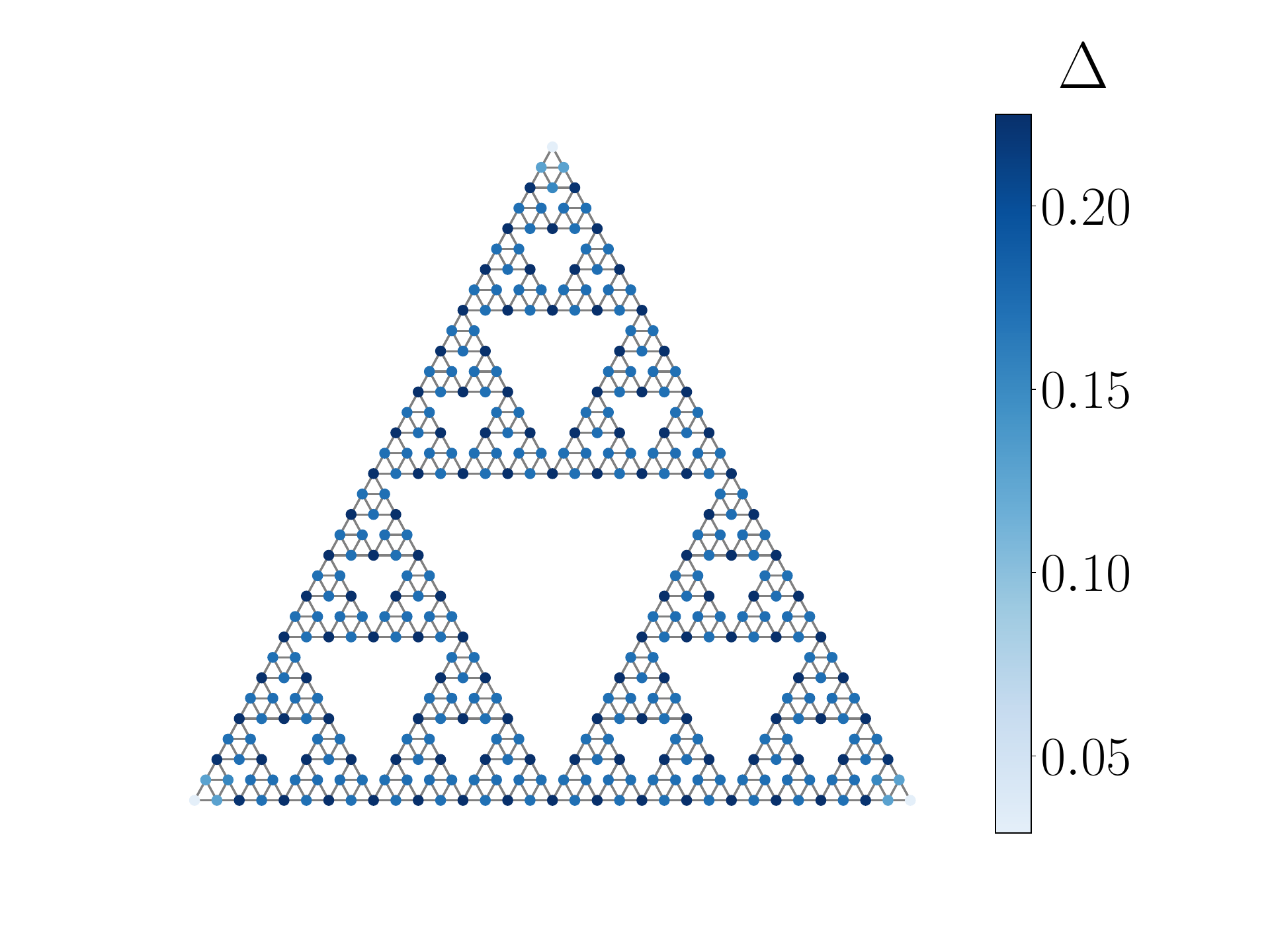}
    \caption{\label{fig:delta_triangle} Cooper pair condensate $\Delta$ at $U=1.0$, chemical potential $\mu=2.0$, temperature $T=0.005$ for the top figures, and $T=0.05$ (around $T_c$ of the undeformed lattice) for the bottom ones. Two structures are shown -- the triangular sample with the disorder ($30\%$ of deleted sites) and the Sierpi\'nski gasket of depth $4$. The sites with $\Delta<0.01$ are indicated by red color.}
\end{figure*}

\begin{figure*}[t!]
    \centering
    \includegraphics[width=0.45\textwidth]{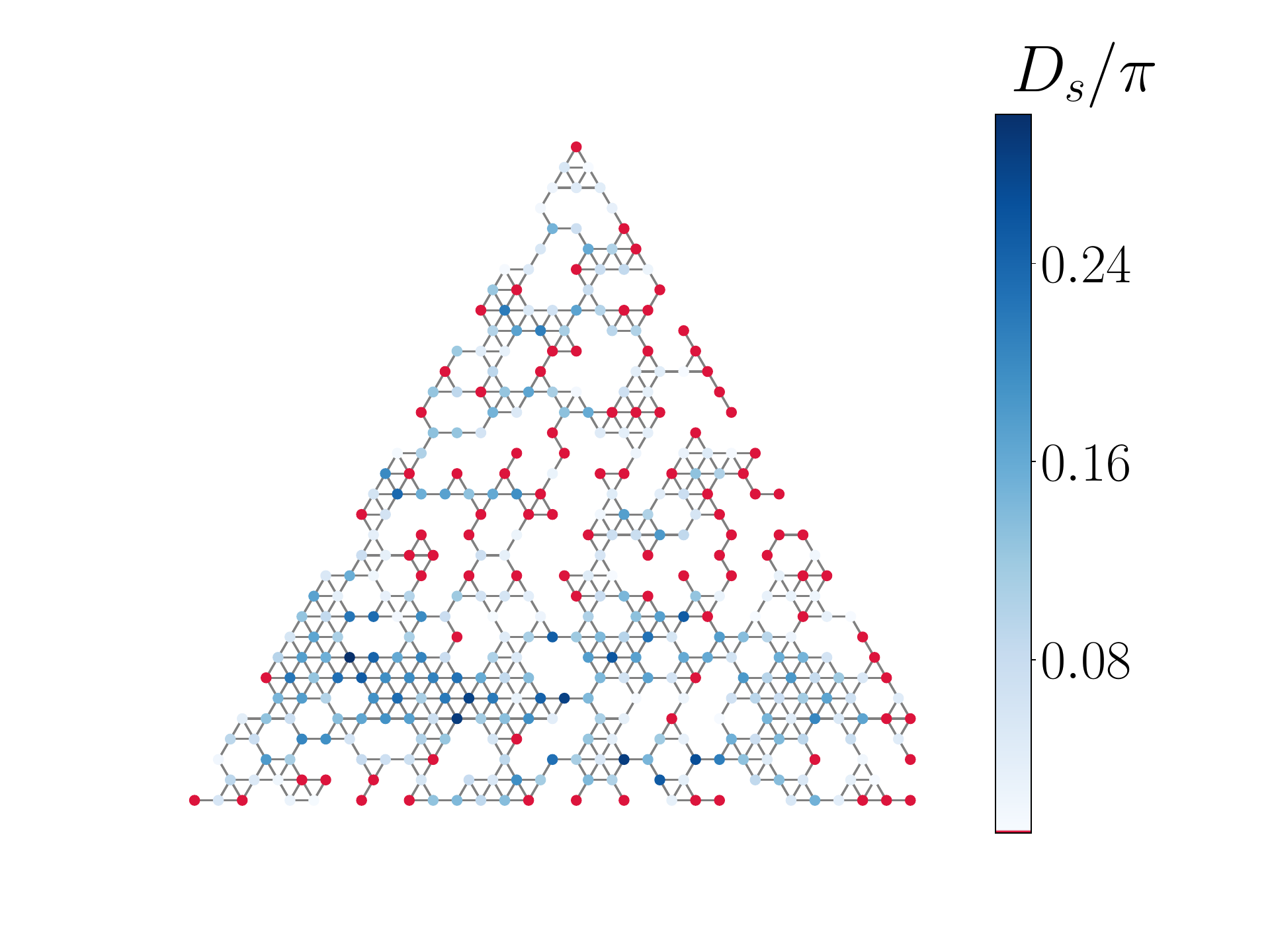}
    \includegraphics[width=0.45\textwidth]{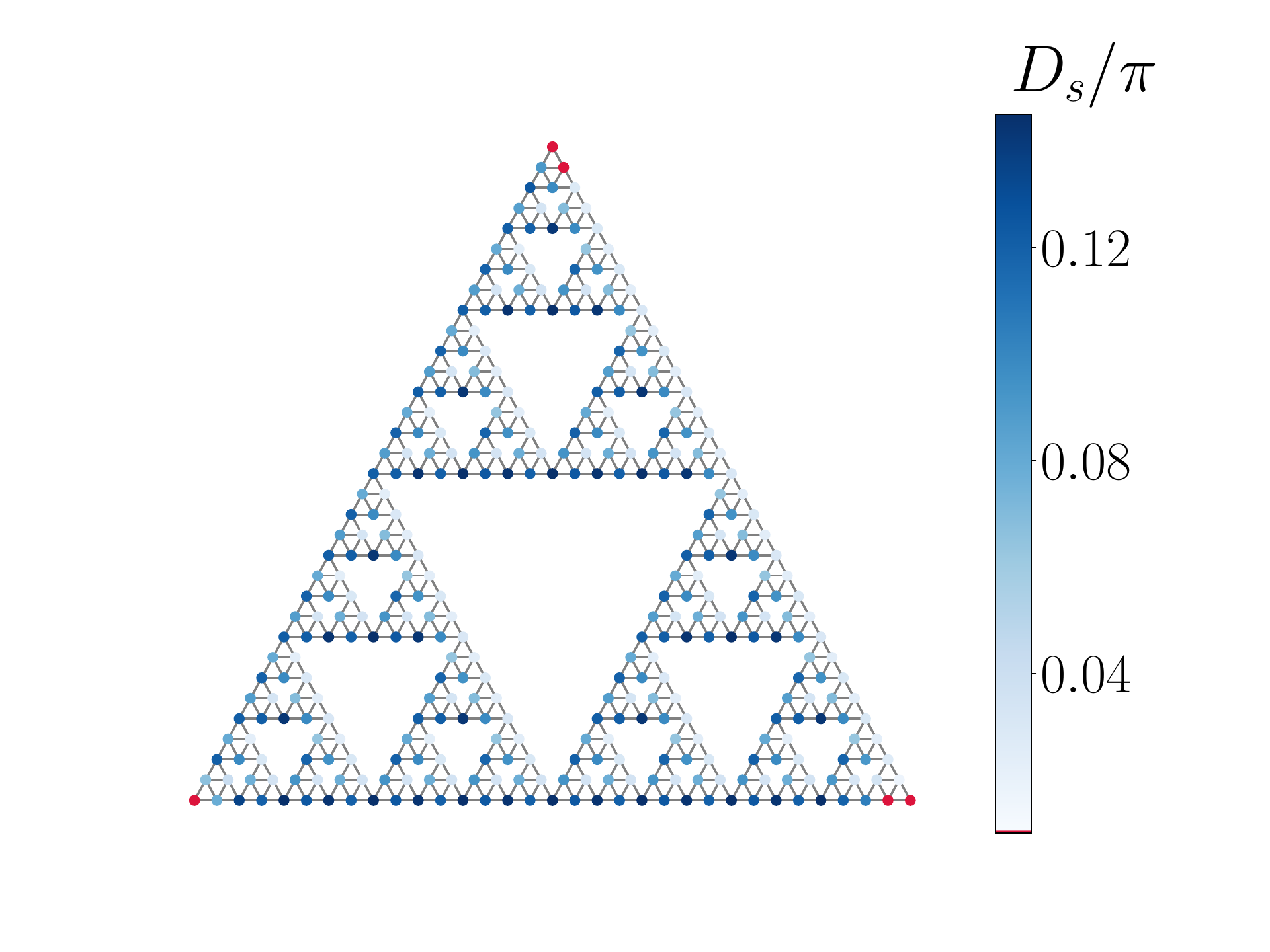}
    \includegraphics[width=0.45\textwidth]{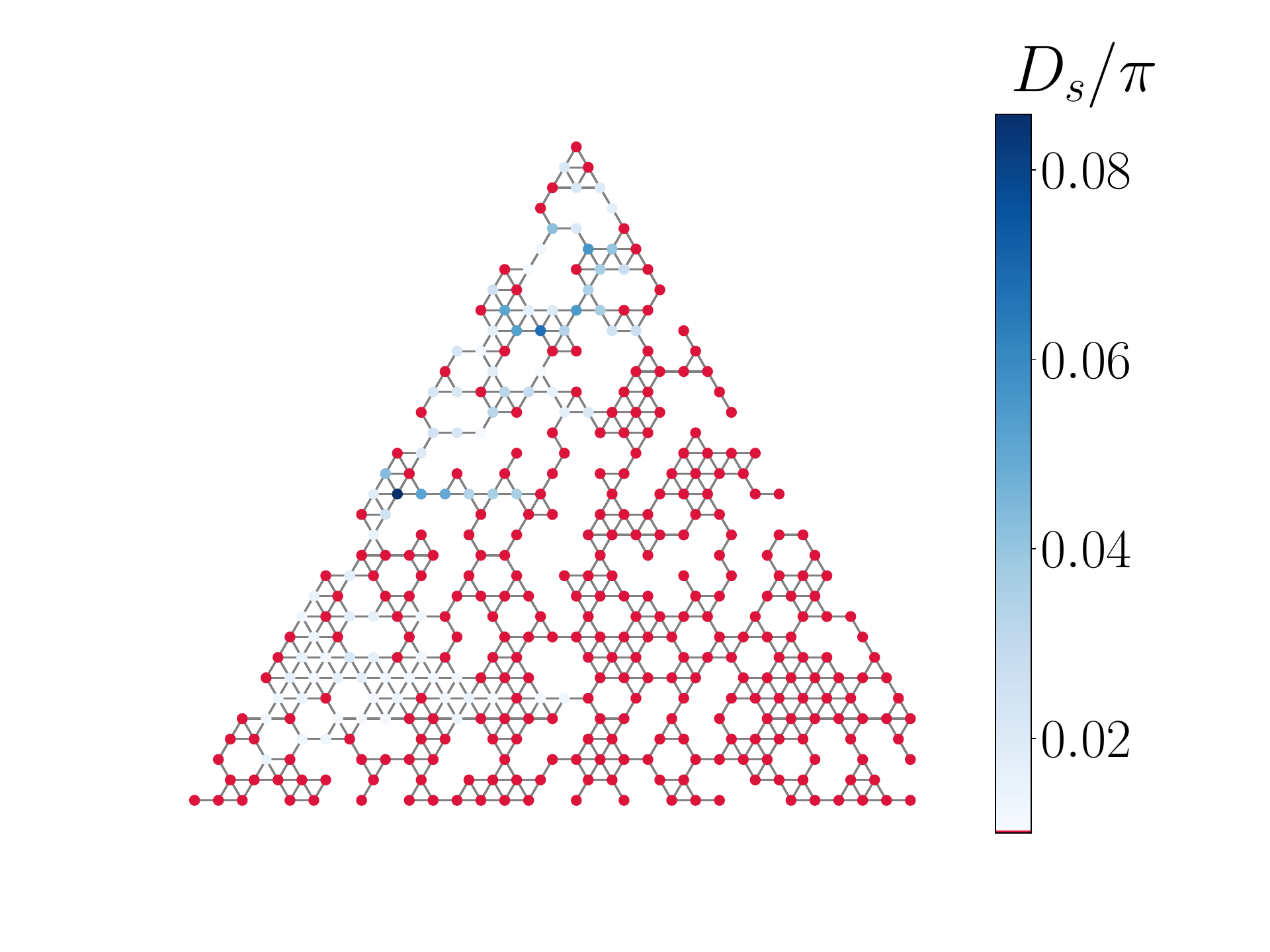}
    \includegraphics[width=0.45\textwidth]{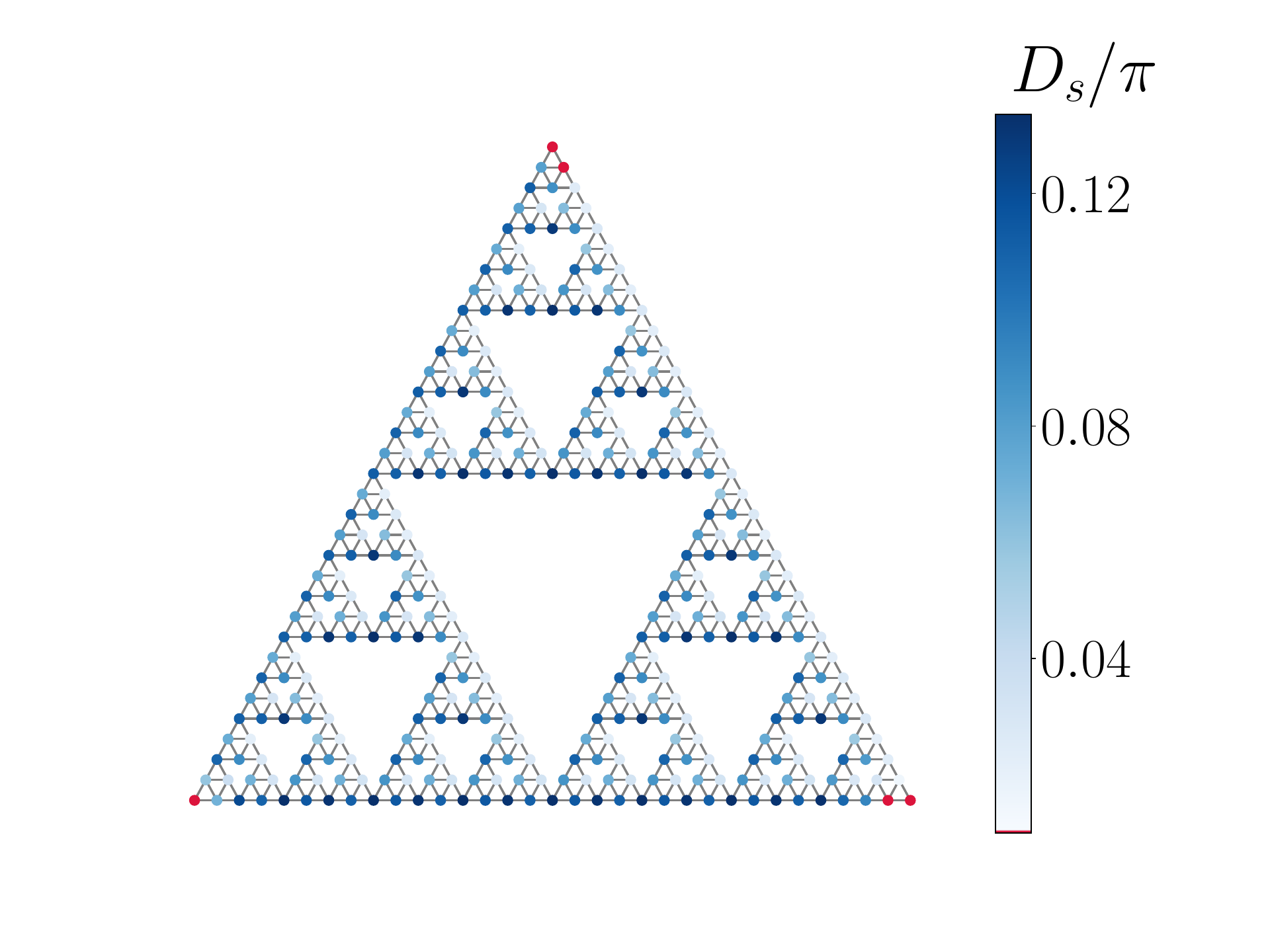}
    \caption{\label{fig:rho_triangle} The superfluid density $D_s/\pi$ at $U=1.0$, chemical potential $\mu=2.0$, temperature $T=0.005$ for the top figures, and $T=0.05$ (around $T_c$ of the undeformed lattice) for the bottom ones. Two structures are shown -- triangular sample with the disorder ($30\%$ of deleted sites) and the Sierpi\'nski gasket of depth $4$. The sites with $D_s/\pi<0.01$ are indicated by red color.}
\end{figure*}
 
Our first goal is to compute the real-space profile of the superconducting condensate across the parametric space of the model. We do that within the self-consistent Bogoliubov-de Gennes approach \cite{BdG1, BdG2, Book}. Within the coherent potential approximation (CPA) for on-site-disordered superconductor, this relatively simple approach \cite{Zittartz1, Zittartz2} was shown to lead to the same results as the complete Eliashberg theory for electron-phonon interactions \cite{MK_Anokhin}. Therefore one can hope that this approach may be sufficient also for our situation.

We apply the mean-field approximation including the Hartree-Fock correction:
\begin{gather}
    n_{i\uparrow} n_{i\downarrow}\approx \langle c_{i\downarrow} c_{i\uparrow} \rangle c^{\dagger}_{i\uparrow} c^{\dagger}_{i\downarrow}+\langle c^{\dagger}_{i\uparrow} c^{\dagger}_{i\downarrow}\rangle c_{i\downarrow} c_{i\uparrow} + \\ \nonumber \langle n_{i\uparrow}\rangle n_{i\downarrow}+ \langle n_{i\downarrow}\rangle n_{i\uparrow}.
\end{gather}
Introducing the local pairing amplitude $\Delta_i$, the mean-field Hamiltonian can be rewritten as follows
\begin{equation}
    H=t\sum_{\langle i,j \rangle\sigma} c^{\dagger}_{i\sigma}c_{j\sigma}-\sum_{i\sigma}\tilde{\mu}_{i,\sigma} n_{i\sigma}+\sum_{i} [\Delta_i c^{\dagger}_{i\uparrow} c^{\dagger}_{i\downarrow}+ \Delta^{*}_i c_{i\downarrow} c_{i\uparrow}],
\end{equation}
with self-consistency conditions given by
\begin{gather}
    \Delta_i=|U|\langle c_{i\uparrow} c_{i\downarrow}\rangle =-|U|\langle c_{i\downarrow} c_{i\uparrow}\rangle\\
    \langle n_{i\sigma}\rangle=\langle c^{\dagger}_{i\sigma} c_{i\sigma}\rangle, \nonumber \\
   \tilde{\mu}_{i,\sigma} = \mu +U \langle n_{i,\sigma}\rangle \nonumber
\end{gather}
To bring the mean-field Hamiltonian to single--particle form, we apply the Bogoliubov transformation:
\begin{gather}
    c^{\dagger}_{i\uparrow}=\sum_{n(E_n \ge 0)}(u^{n*}_{i\uparrow}\gamma^{\dagger}_{n\uparrow}-v^{n}_{i\uparrow}\gamma_{n\downarrow}), \\
    c^{\dagger}_{i\downarrow}=\sum_{n(E_n \ge 0)}(u^{n*}_{i\downarrow}\gamma^{\dagger}_{n\downarrow}+v^{n}_{i\downarrow}\gamma_{n\uparrow}) \nonumber
\end{gather}
where $\gamma_n$ and $\gamma^{\dagger}$ are quasiparticle operators satisfying the fermionic anti-commutation relations and the relation:
\begin{gather}
    \langle\gamma^{\dagger}_{n\sigma} \gamma_{n\sigma}\rangle= \frac{1}{\exp(E_{n\sigma}/k_B T)+1}.
\end{gather}
The Fermi-Dirac distribution function is denoted as $f(E_{n\sigma})$ in what follows \footnote{The chemical potential $\mu$ is included via the diagonal term in the Hubbard model Eq. \eqref{eq:Hubbard_model}}.

In the new basis, the eigenvalue equation acquires the form
\begin{gather}
    \begin{pmatrix}
    \hat K&\hat\Delta\\\hat\Delta^{*}&-\hat K
    \end{pmatrix}    \begin{pmatrix}
    u^{n}_{\uparrow}\\v^{n}_{\downarrow}
    \end{pmatrix}=E_{n\uparrow}\begin{pmatrix}
    u^{n}_{\uparrow}\\v^{n}_{\downarrow}
    \end{pmatrix}, \label{eq:Gorkov}
\end{gather}
where $\hat K$ is the matrix representing the free part of the Hamiltonian, and $\hat \Delta=\mbox{diag}(\Delta_1,\dots\Delta_N)$ with $N$ being the total number of lattice sites. Due to the absence of spin-orbit coupling and real-valuedness of the eigenfunctions, the other vector $(u_{n\downarrow}, v_{n\uparrow})^{T}$ satisfies exactly the same equation. Hence, we need to solve the equation only for one of them, and from now on we omit the spin indices.

The self-consistency conditions are then:
\begin{gather}
    \Delta_i=\frac{|U|}{2}\sum_{n}u^{n}_i v^{n*}_i \tanh(\frac{E_n}{2k_B T}), \label{eq:selfcons_delta}\\
    \langle n_{i}\rangle=\sum_{n}[|u^n_i|^2 f(E_n)+|v^n_i|^2 f(-E_n)], \label{eq:selfcons_density}
\end{gather}
where the sums run over both the positive and the negative parts of the spectrum, and we solve Eqs. \eqref{eq:Gorkov}-\eqref{eq:selfcons_density} iteratively. 

To unveil the effects of fractal geometry on superconductivity, for each point $(T, \mu, U)$ in the space of parameters, we solve the equations for three cases. First, for a regular lattice (square or triangular) with large linear size $L=1000$ ($N=L^2$) and periodic boundary conditions, one reconstructs the phase diagram in the thermodynamic limit. Secondly, for a regular lattice, but with a smaller linear size $L=27$ and open boundary conditions -- to see how the finite-size effects affect the phase diagram. Finally, we solve the equations for the fractal structures with side length $L=27$ (the Sierpi\'nski carpet in the case of the parental square lattice, and the gasket for the parental triangular lattice). The results for $U=1.0$ are shown in Fig. \ref{fig:phase_diagram}. From these pictures, it is clear that fractal geometry brings more than just finite-size effects to the table. In the case of the Sierpi\'nski carpet, it broadens the superconducting dome and makes its structure highly irregular, but has little effect on the optimal critical temperature, even lowering it a bit. On the contrary, the gasket geometry leads to a strong increase of maximal $T_c$, doubling it for the chosen value of attractive coupling \cite{Babaev} (in Supplemental Note 2, we provide results for other values of $U$). One-dimensional sections of the phase diagrams at optimal values of chemical potential are shown in Fig.\ref{fig:1d_slice}, and the spatial profile of condensate in Fig.\ref{fig:delta_triangle} (for the carpet, it is provided in Supplemental Note 3). Visually, the condensate seems to be suppressed in the corners as compared to other points of the lattice, but in fact still has non-negligible value of $\Delta\sim 0.03$.

To better understand the reason behind the critical temperature elevation, in Supplemental Note 4, we compute the relation between $T_c$ and the density of states at the Fermi level $g(E_F)$. In particular, it is shown that for the Sierpinski gasket, the relation $T_c \simeq \theta e^{-\alpha/Ug(E_F)}$ is satisfied with the only difference that $\alpha\simeq 0.6$ instead of $\alpha=1$ as follows from the conventional BCS theory, and the density of states acquire larger values than for the regular triangular lattice. Both these aspects have a positive effect on $T_c$.

However, the enhancement of critical temperature does not automatically imply that superconductivity survives at higher temperatures as a global phenomenon. It is still possible that fractal geometry makes individual Cooper pairs more robust to thermal fluctuations, but breaks the long-range coherence of the global superconducting condensate and hence prohibits the large-scale supercurrent flow through the sample (a possible scenario for disordered superconductors \cite{disorder_kill_SC, Feigelman}). In the bottom panel of Fig.\ref{fig:delta_triangle}, we see that non-zero condensate $\Delta$ survives across the lattice for both the disordered and the fractal systems at the temperature that would be critical for the undeformed sample, indicating an increase in $T_c$ in both cases. The long-range coherence can be analyzed by computing the superfluid stiffness of the system in the presence of an external electric field. Then, the condition for the existence of a supercurrent flow between two points of the sample is the existence of a line connecting them along which the superfluid stiffness is non-zero everywhere.

Assume that the electric field $E_x$ is applied along the $x$-axis (rightwards along the lower leg of the triangle in Figs. \ref{fig:delta_triangle}, \ref{fig:rho_triangle}). The superfluid stiffness is then given by \cite{Scalapino}:
\begin{equation}
    \frac{D_s}{\pi}=\Pi_{xx}(q_x=0,q_y\to0,\omega=0)-\langle K_x \rangle, \label{eq:stiffness_main}
\end{equation}
where $\Pi_{xx}$ is the retarded current-current correlation function, $\omega$ and ${\bf q}$ are the frequency and the wave vector of the applied electric field correspondingly, and $\langle K_x \rangle$ is the kinetic energy density of electrons moving in the direction of the applied field. Explicit form of the r.h.s. of \eqref{eq:stiffness_main} used for numerical computations is given in Supplemental Note 1.

We computed the on-site superfluid stiffness $D_s/\pi$ for the Sierpi\'nski gasket and for a disordered sample of the triangular lattice with $\sim 30\%$ of sites removed (so that both lattices have approximately the same number of sites). The results are shown in Fig. \ref{fig:rho_triangle}. As one can see, for the disordered sample, the profile of $D_s/\pi$ is highly irregular, and the stiffness vanishes in many points splitting the system into a number of disconnected superconducting islands and making the global flow of supercurrent through the sample impossible. On the other hand, for the Sierpi\'nski gasket, $D_s/\pi$ is below threshold value of $0.01$ only on the corner sites where the current induced by $E_x$ terminates, leaving the rest of the lattice a connected component capable of hosting a global current. The case of Sierpi\'nski carpet is shown in Supplemental Note 3.

Our considerations demonstrate, on the proof-of-concept level, that a certain fractal geometry can lead to a considerable enhancement of superconductivity as compared to the original parental crystalline lattice. Here, we have analyzed it within a minimal framework, and several further steps are required to gain an in-depth understanding of this phenomenon. For example, it should be noted that the effect of $T_c$ elevation for boundary states in regular crystalline superconductors has been recently discussed \cite{Babaev_boundary}. Since fractal lattices have boundaries ``everywhere'' in the bulk, it is possible that similar effect plays a certain role here. When discussing superconductivity in fractals, it is tempting to compare fractals with their cousin geometry -- quasicrystals. Just like fractals, quasicrystals exhibit discrete scale invariance, and superconductivity in such systems has been observed \cite{quasicrystals} and theoretically shown to be strongly affected by the unconventional structure of the lattice \cite{Sakai}. However, they are not characterized by interpenetration of boundary and bulk, and do not possess non-trivial ramification properties. Given that enhancement of $T_c$ is observed for the finitely ramified gasket fractal, and not for the infinitely ramified carpet, connectivity properties are of critical importance in this context, and geometry effects on superconductivity in fractals and quasicrystals are fundamentally different.

In this paper, the attractive Hubbard model was analyzed within the self-consistent Hartree-Fock approximation that takes into account anomalous averages and charge amplitudes. Potentially, self-consistent treatment of spin amplitudes can also lead to important magnetic effects that can alter the phase diagram of the model, such as competition of superconducting and magnetic ordering, emergence of magnons, and renormalization of the coupling constant due to soft magnetic modes (spin waves). These phenomena must be studied in detail, and we are planning to address them in the subsequent work.

Also, here we studied only the $s$-wave superconductivity, 
while superconducting condensates with other types of symmetry, such as $p$- or $d$-wave, can be affected by fractal geometries in different ways. This can be analyzed on the BdG mean-field level as well but in the extended attractive Hubbard model. The next natural steps will be to develop a complete Eliashberg theory of superconductivity in fractal geometries, and to employ more sophisticated real-space methods (real-space constrained random phase approximation \cite{real_space} or variational finite-temperature Monte Carlo algorithms \cite{VMC1, VMC2}) to study fractal superconductors in the regime of strong repulsive interactions in more realistic models such as the $t-t'$ repulsive Hubbard model \cite{t-t-prime1,t-t-prime2,t-t-prime3} and to account for the long-range Coulomb interactions \cite{Coulomb1,Coulomb2,Coulomb3}. Should this phenomenon emerge in other regimes apart from those considered in this paper, it could define a new way to search for novel high-Tc superconductors based on engineered atomic deformations of lattice geometries of the natural crystalline superconductors.

\acknowledgments

We thank D.~Ageev for fruitful and extensive discussions on many related topics, and acknowledge discussions with O.~Eriksson and E.~Stepanov. The work of A.A.B., A.A.I., and M.I.K. was supported by the European Research
Council (ERC) under the European Union’s Horizon 2020 research and innovation program, grant agreement 854843-
FASTCORR. A.A.B. and M.I.K. acknowledges the research program “Materials for the Quantum Age” (QuMat) for financial support. This program (registration number 024.005.006) is part of the Gravitation program financed by the Dutch Ministry of Education, Culture and Science (OCW). The data that support the findings of this
study are available from A.A.I. or A.A.B. upon reasonable request.


\begin{thebibliography}{100}
\bibitem{Babaev} Similar results have been obtained by E. Babaev and collaborators (KTH Stockholm), though were not published.
\end{thebibliography}
\end{document}


\renewcommand{\figurename}{Supplementary Figure}

\begin{center}{\Large \textbf{%
    Supplementary Information to the paper ``Strong enhancement of superconductivity on finitely ramified fractal lattices''
}}\end{center}

\begin{center}
Askar A. Iliasov,
Robert Canyellas,
Mikhail I. Katsnelson,
Andrey A. Bagrov
\end{center}

\section*{Supplementary Note 1: Derivation of the superfluid stiffness}
\label{sec:stiffness}
Kinetic energy (Eq. 10 of the main text) is defined by the following expression:
\begin{equation}
    \langle K_x \rangle=\frac{1}{N}\sum_{l,\sigma} K_{l,\sigma,x}=-\frac{t}{N} \langle \sum_{l\sigma}\sum_{m}(x_l-x_m)^2[c^{\dagger}_{k\sigma}c_{m\sigma}+c^{\dagger}_{l\sigma}c_{m\sigma} ]\rangle,
\end{equation}
where $N$ is the number of lattice sites, we set the lattice constant $a=1$, and the summation over index $m$ goes over the neighbors of the site $l$ with larger $x$ coordinate, $x_m>x_l$.

The retarded current-current correlator can be obtained by the analytical continuation from the expression written in the Matsubara frequencies, $\omega_n\to\omega+i\delta$:
\begin{equation}
    \Pi_{xx}({\bf q},\omega_n)=\frac{1}{N}\int^{\beta}_{0} e^{i\omega_n \tau}\langle j_x({\bf q},\tau)j_x(-{\bf q},0) \rangle,
\end{equation}
where the current operator $j_x$ is defined as:
\begin{equation}
j_x=it\sum_{l\sigma}\sum_m e^{-i{\bf q}\cdot{\bf r_l}}(x_m-x_l)[c^{\dagger}_{m\sigma}c_{l\sigma}-c^{\dagger}_{l\sigma}c_{m\sigma} ],
\end{equation}
with $\bf{r_l}$ being the Cartesian coordinates of the lattice sites.

After the Bogoliubov transformation, these terms acquire the following forms.

The kinetic energy expectation value:
\begin{equation}
    \langle K_x \rangle=-\frac{2t}{N}\sum_{n,l}\sum_m (x_m-x_l)^2[(u^{n*}_{m}u^n_l+u^{n*}_{l}u^n_{m})f(E_n)+(v^{n*}_{m}v^n_l+v^{n*}_{l}v^n_{m})f(-E_n)]. \label{eq:kinetic_global}
\end{equation}

The current-current correlator:
\begin{equation}\label{eq:Pixx}
    \Pi_{xx}({\bf q},\omega_n)=\frac{1}{N}\sum_{n_1,n_2}\frac{A_{n_1,n_2}({\bf q})[A^{*}_{n_1,n_2}({\bf q})+D_{n_1,n_2}(-{\bf q})]}{\omega+(E_{n_1}-E_{n_2})+i\delta}[f(E_{n_1})-f(E_{n_2})],
\end{equation}
where $A_{n1,n_2}({\bf q})$ and $D_{n1,n_2}({\bf q})$ are defined as:
\begin{gather}
    A_{n_1,n_2}({\bf q})=2\sum_{l} e^{-i {\bf q}\cdot {\bf r}_l}\sum_m (x_m - x_l)[u^{n_1 *}_{m}u^{n_2}_l-u^{n_1*}_{l}u^{n_2}_{m}]\\
    D_{n_1,n_2}({\bf q})=2\sum_{l} e^{-i {\bf q}\cdot {\bf r}_l}\sum_m (x_m - x_l)[v^{n_1 *}_{m}v^{n_2}_l-v^{n_1*}_{l}v^{n_2}_{m}].
\end{gather}

\begin{figure*}[t!]
    \centering
    \includegraphics[width=0.95\textwidth]{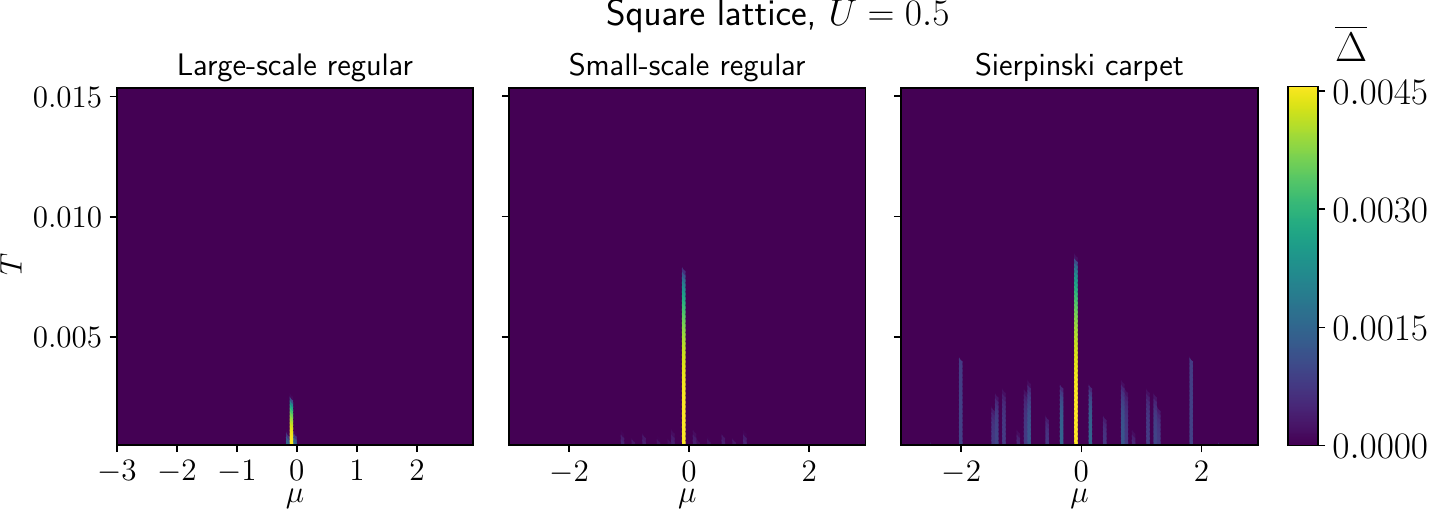}
    \vskip 15pt
    \includegraphics[width=0.95\textwidth]{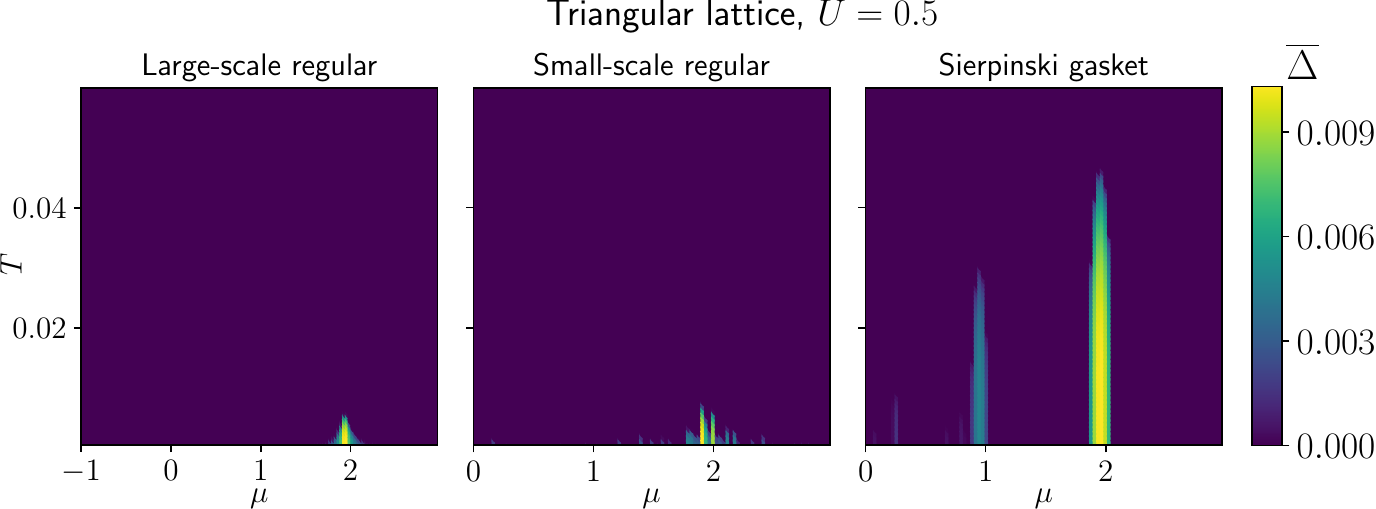}
    \caption{\label{fig:phase_diagram_0.5}$T-\mu$ phase diagrams of the attractive Hubbard model on regular lattices and fractal structures at $U=0.5$. The description given in the caption of Fig. 1 of the main text applies here.}
\end{figure*}

To compute the superfluid stiffness, we first set $\omega=0$ in the Eq. \eqref{eq:Pixx}, and then take the limit ${\bf q}\to 0$ obtaining:
\begin{equation}
    \Pi_{xx}({\bf q}\to0,\omega_n)=\frac{1}{N}\sum_{n_1,n_2}\frac{f(E_{n_1})-f(E_{n_2})}{E_{n_1}-E_{n_2}}A_{n_1,n_2}({\bf q}\to0)[A^{*}_{n_1,n_2}({\bf q}\to0)+D_{n_1,n_2}({\bf q}\to0)] \label{eq:jj_global}
\end{equation}
where, in the case of $E_{n_1}=E_{n_2}$, the prefactor should be calculated as $\partial_E f(E)$.

The stiffness computed from Eqs. \eqref{eq:kinetic_global} and \eqref{eq:jj_global} is averaged over the lattice sites. To calculate its spatial profile across the lattice, we need to avoid averaging over the sites  by omitting the summations over $l$ in Eq. \eqref{eq:kinetic_global} and in the first $A_{n_1,n_2}({\bf q})$ multiplier in the numerator of Eq. \eqref{eq:jj_global}, as well as the normalizing $1/N$ factors. The local value of $\Pi_{xx}$ on site $l$ is given by:
\begin{equation}
    \Pi_{xx}({\bf q},\omega_n,l)=\sum_{n_1,n_2}\frac{A'_{n_1,n_2}({\bf q},l)[A^{*}_{n_1,n_2}({\bf q})+D_{n_1,n_2}(-{\bf q})]}{\omega+(E_{n_1}-E_{n_2})+i\delta}[f(E_{n_1})-f(E_{n_2})]
\end{equation}
with
\begin{equation}
A'_{n_1,n_2}({\bf q})=2 e^{-i {\bf q}\cdot {\bf r}_l}\sum_m (x_m - x_l)[u^{n_1 *}_{m}u^{n_2}_l-u^{n_1*}_{l}u^{n_2}_{m}]
\end{equation}
\section*{Supplementary Note 2: Phase diagrams at diverse U}
\label{sec:other_U}
\begin{figure*}[t!]
    \centering
    \includegraphics[width=0.95\textwidth]{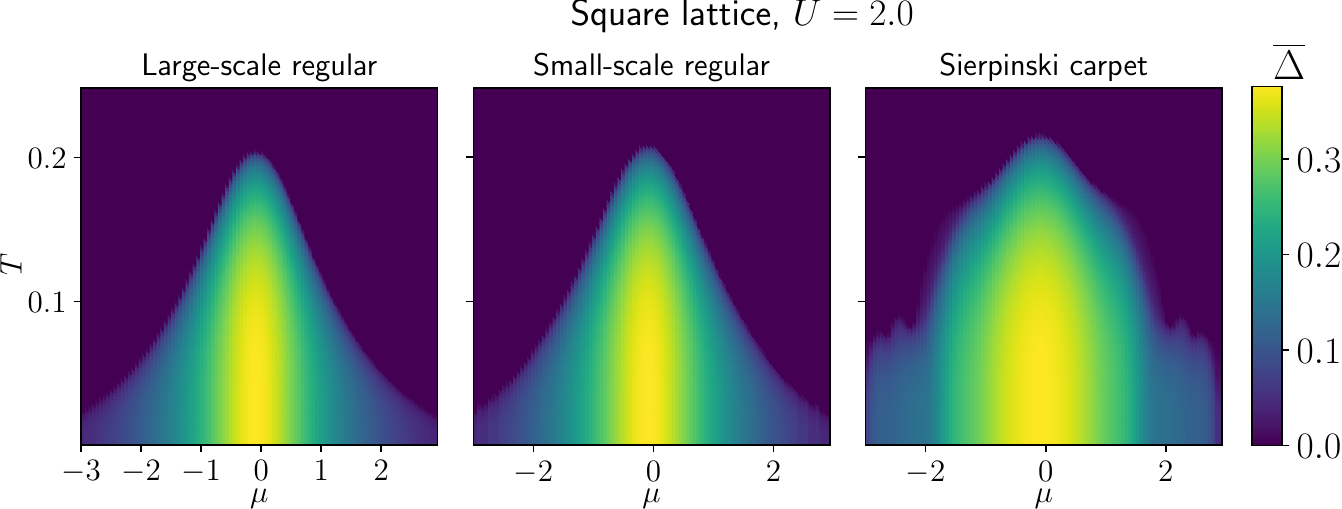}
    \vskip 15pt
    \includegraphics[width=0.95\textwidth]{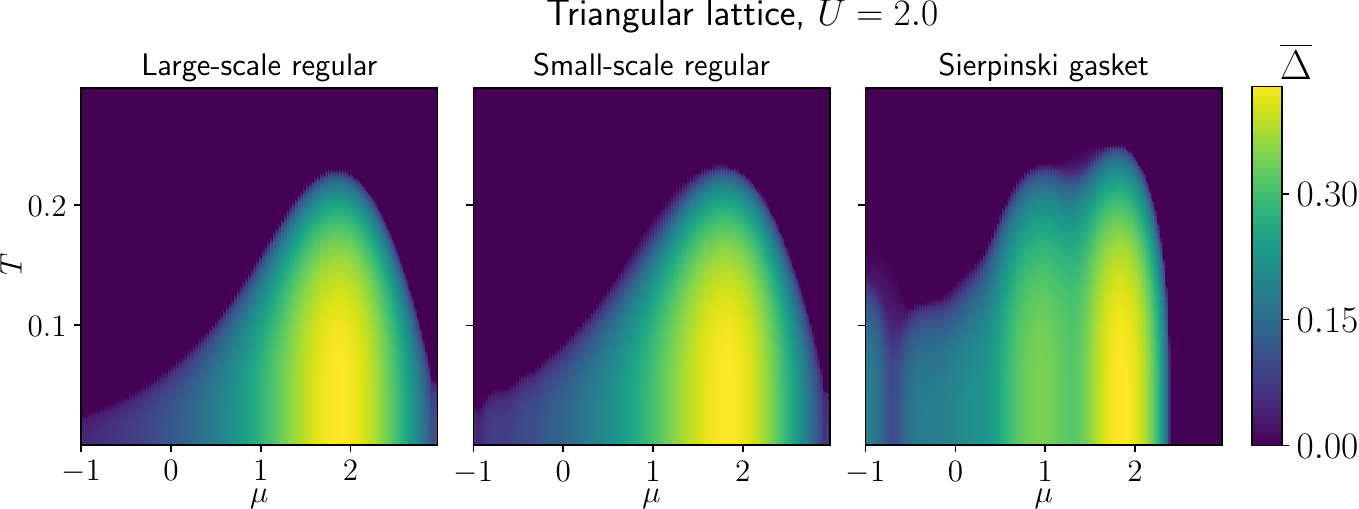}
    \caption{\label{fig:phase_diagram_2.0}$T-\mu$ phase diagrams of the attractive Hubbard model on regular lattices and fractal structures at $U=2.0$. The description given in the caption of Fig. 1 of the main text applies here.}
\end{figure*}
In the main text, we took $U=1.0$, but it is also interesting to consider how fractal geometry affects $T_c$ in the regimes of stronger or weaker attraction. For $U=0.5$ and $U=2.0$ without the Hartree-Fock corrections, the phase diagrams are shown in Figs.\ref{fig:phase_diagram_0.5}, \ref{fig:phase_diagram_2.0}. From this figures, it is clear that, at lower values of $U$, where pairing is weak and $T_c$ is low, the gasket-type fractality leads to much stronger {\emph{relative}} increase in $T_c$, while in the regime of very strong attraction this effect becomes negligible. One can speculate that it can be related to the smearing of the Cooper pairs over the fractal geometry. When attraction is weak and the coherence length is large, the Cooper pair embraces a larger part of the lattice, and there are a few iterations of the fractal inside the pair, alternating its properties. On the other hand, at strong attraction, the Cooper pair is compact and the role played by fractality reduces to local boundary effects.

\section*{Supplementary Note 3: Spatial $\Delta$ and $D_s$ profiles on the Sierpi\'nski carpet}
\begin{figure*}[t!]
    \centering
    \includegraphics[width=0.45\textwidth]{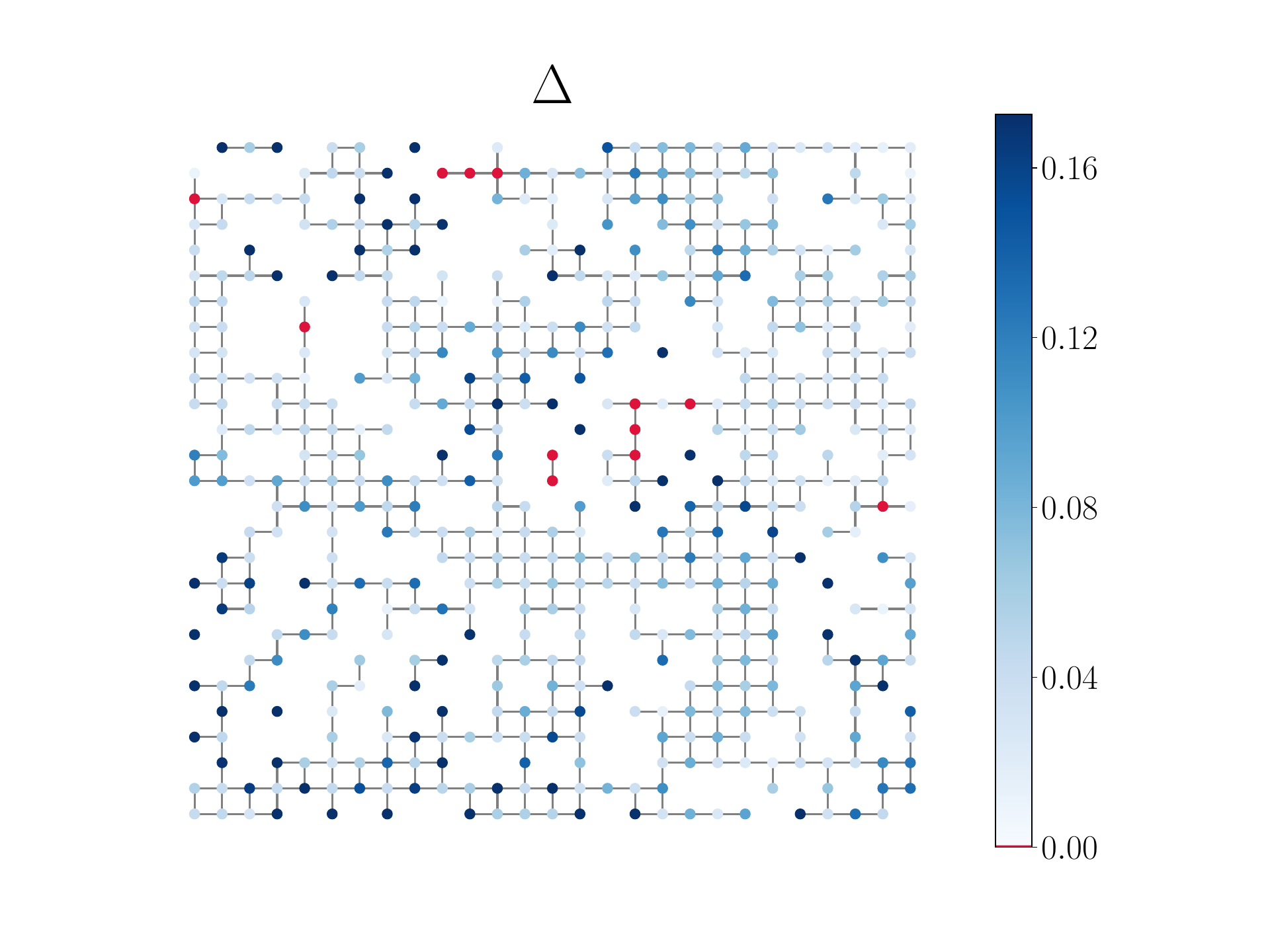}
    \includegraphics[width=0.45\textwidth]{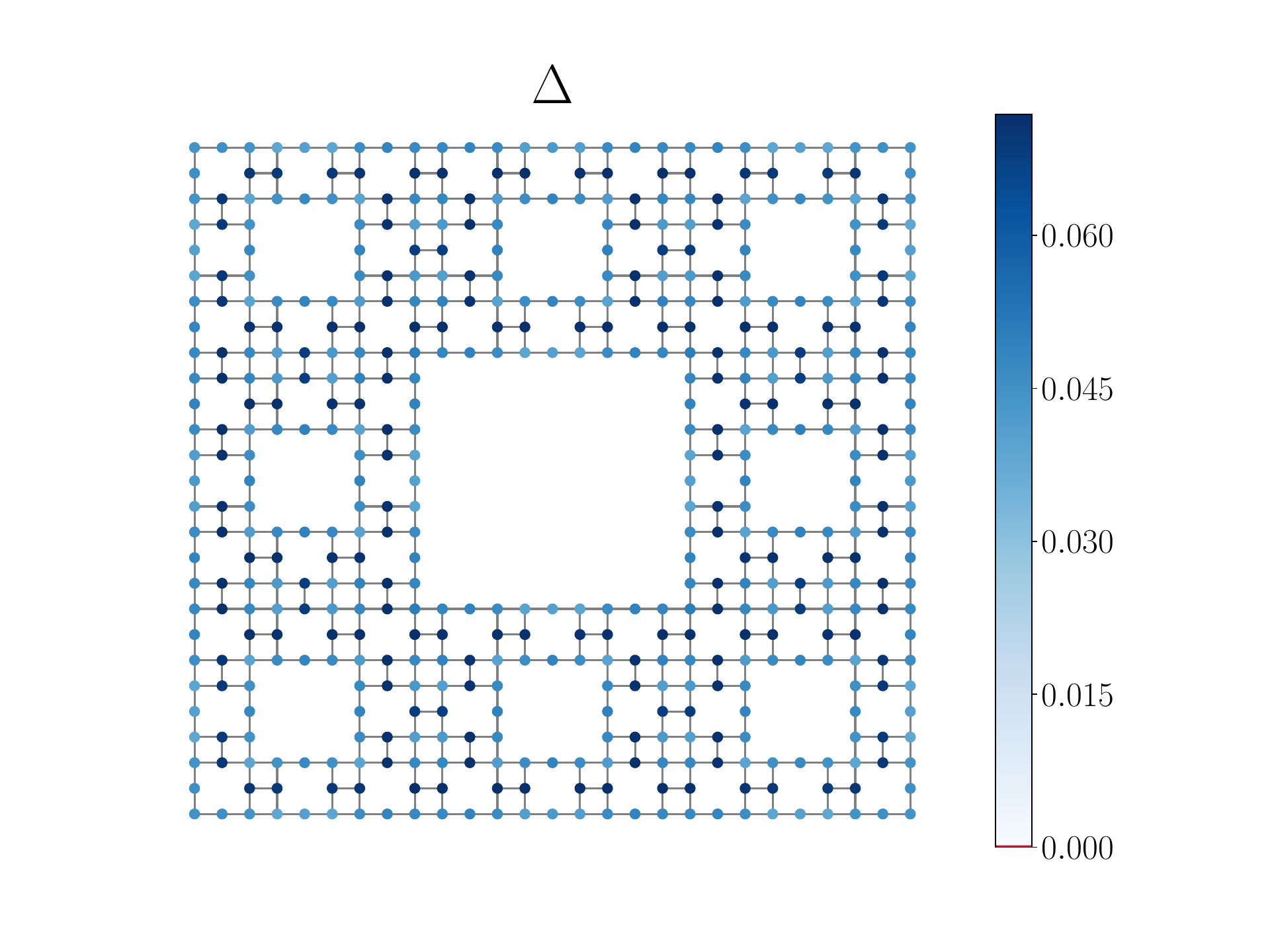}
    \includegraphics[width=0.45\textwidth]{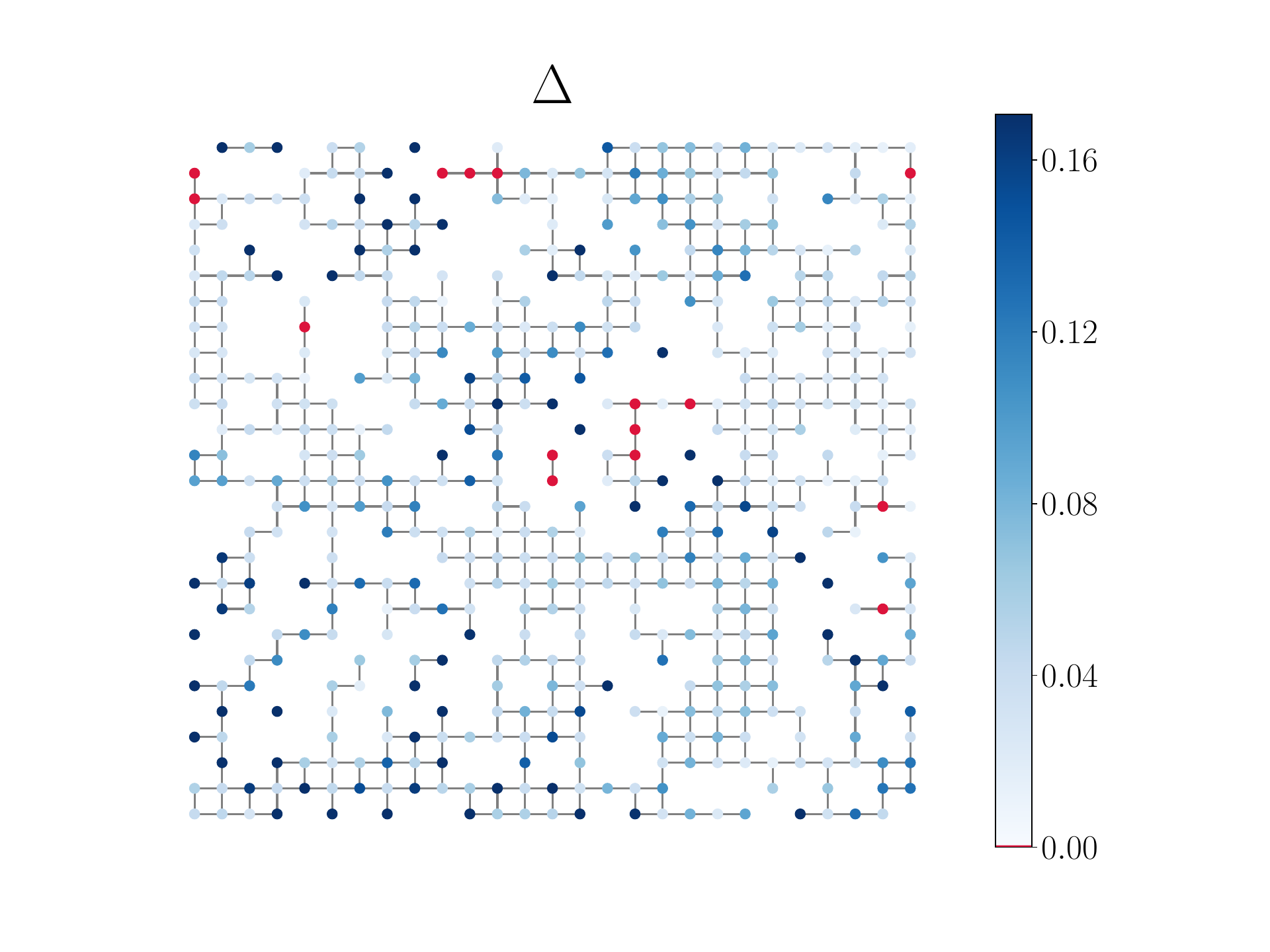}
    \includegraphics[width=0.45\textwidth]{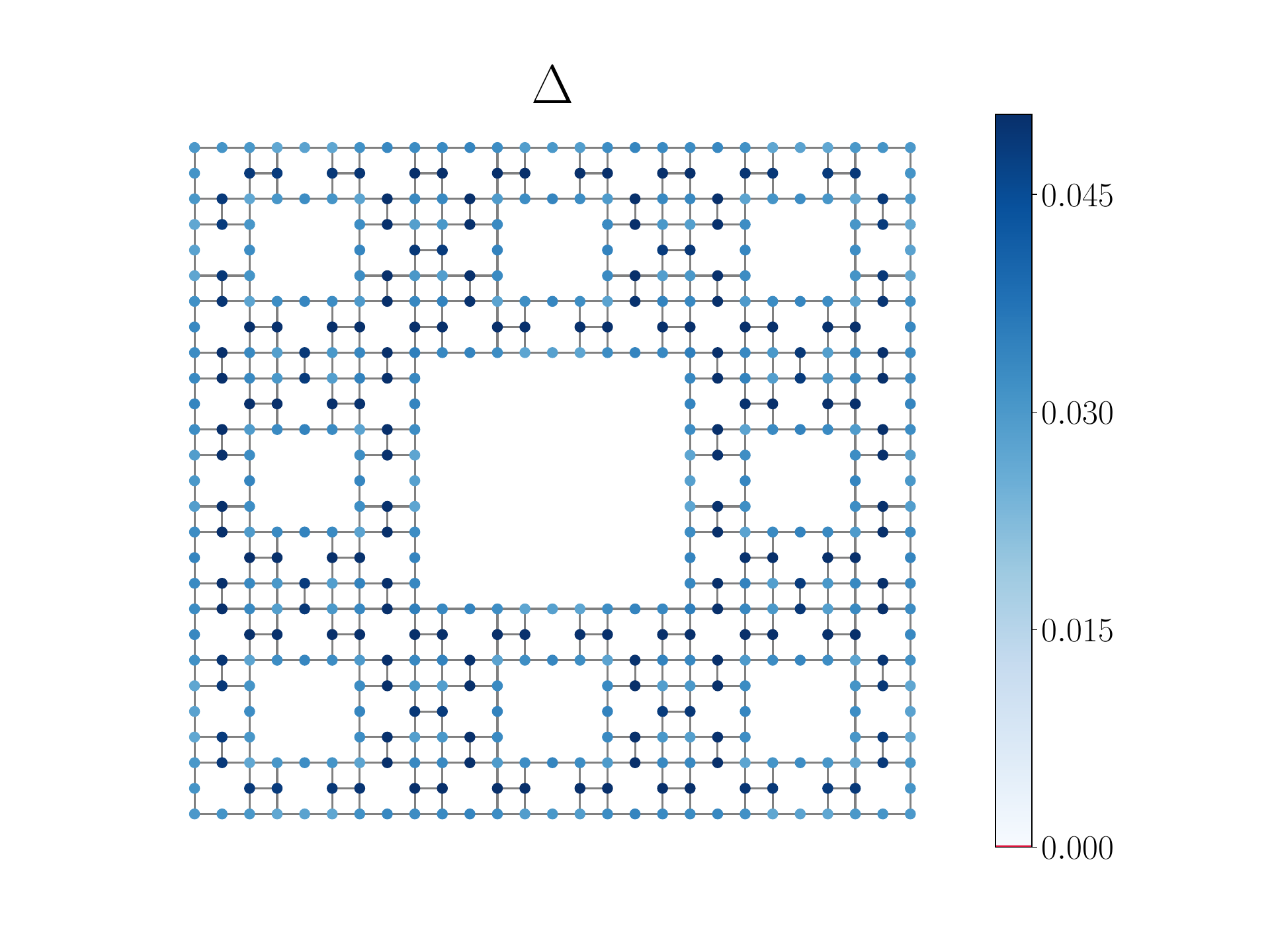}
    \caption{\label{fig:delta_square} Cooper pair condensate $\Delta$ at $U=1.0$, chemical potential $\mu=0.0$, temperature $T=0.005$ for the top figures, and $T=0.025$ (around $T_c$ of the undeformed lattice) for the bottom ones. Two structures are shown -- square sample with the disorder ($30\%$ of deleted sites) and the Sierpi\'nski carpet of depth $3$. The sites with $\Delta<0.01$ are indicated by red color.}
\end{figure*}

\begin{figure*}[t!]
    \centering
    \includegraphics[width=0.45\textwidth]{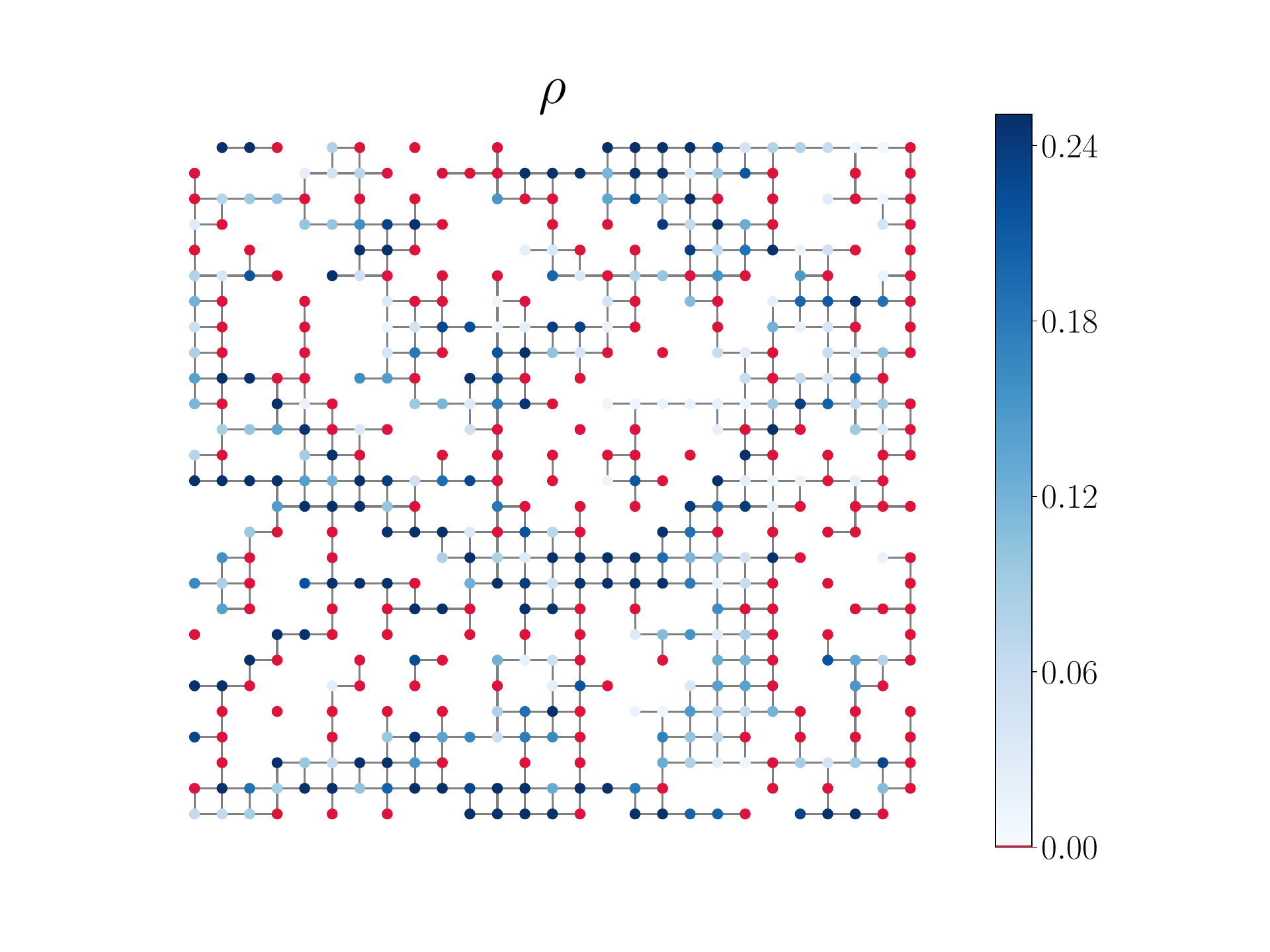}
    \includegraphics[width=0.45\textwidth]{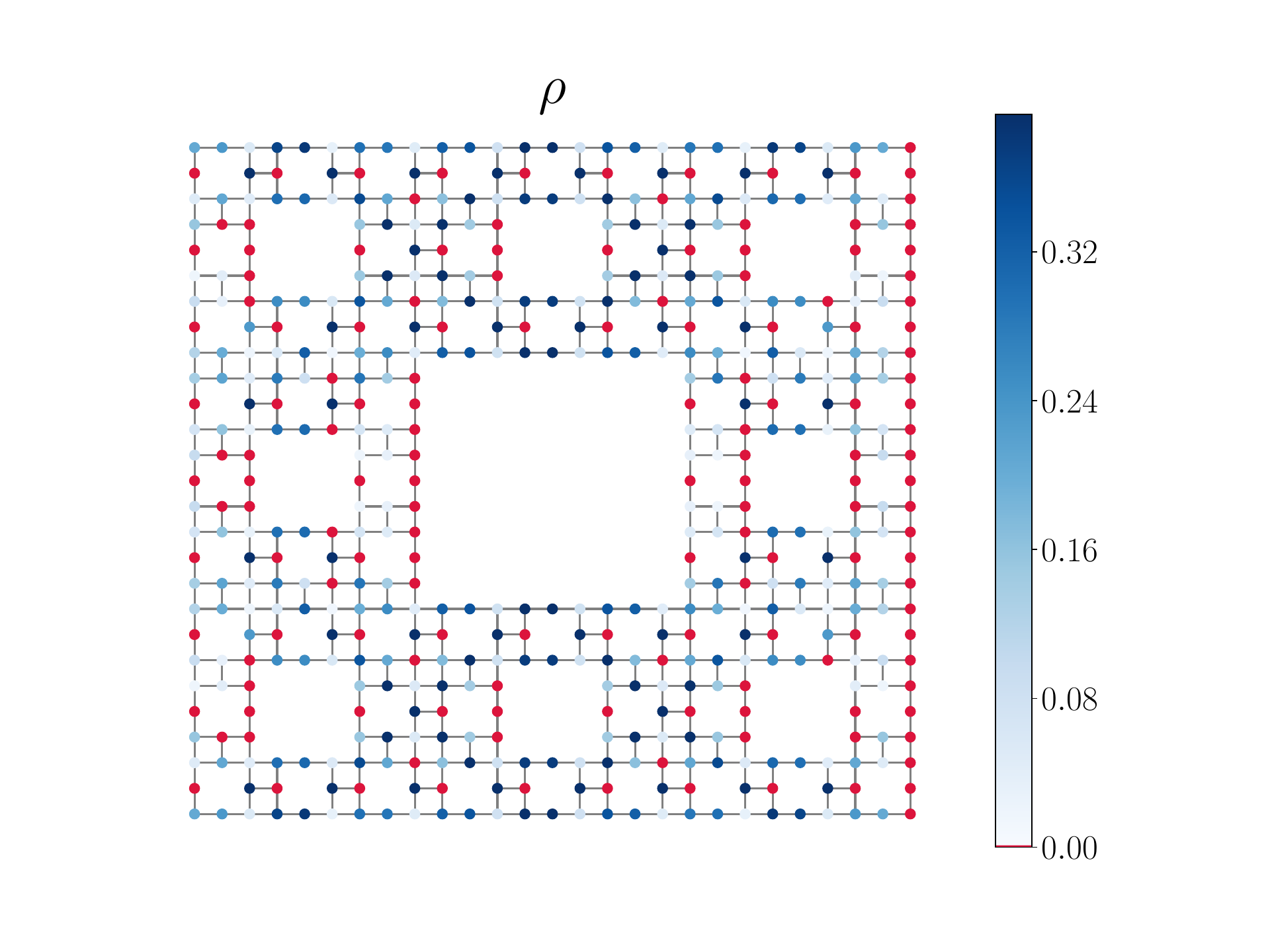}
    \includegraphics[width=0.45\textwidth]{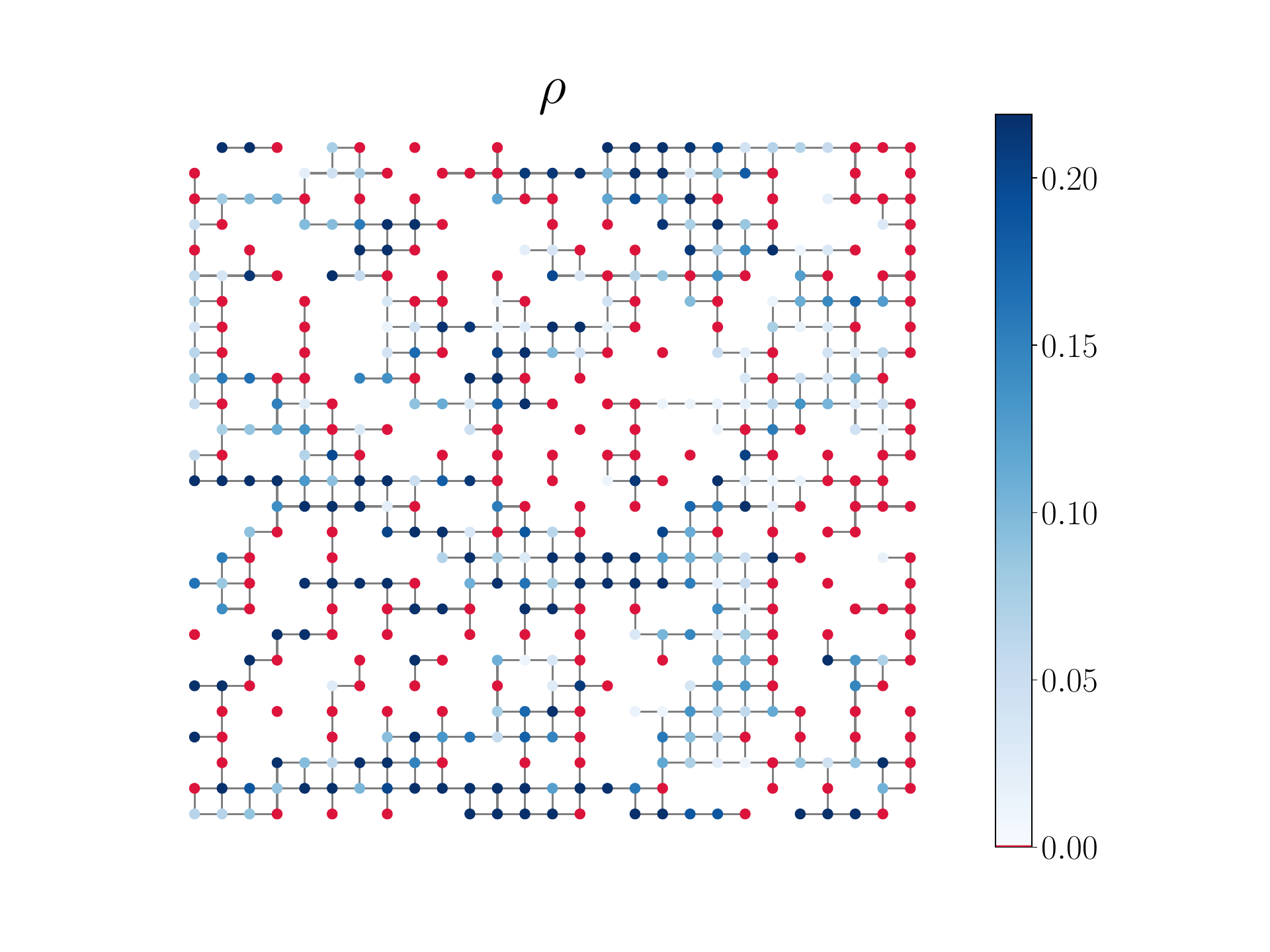}
    \includegraphics[width=0.45\textwidth]{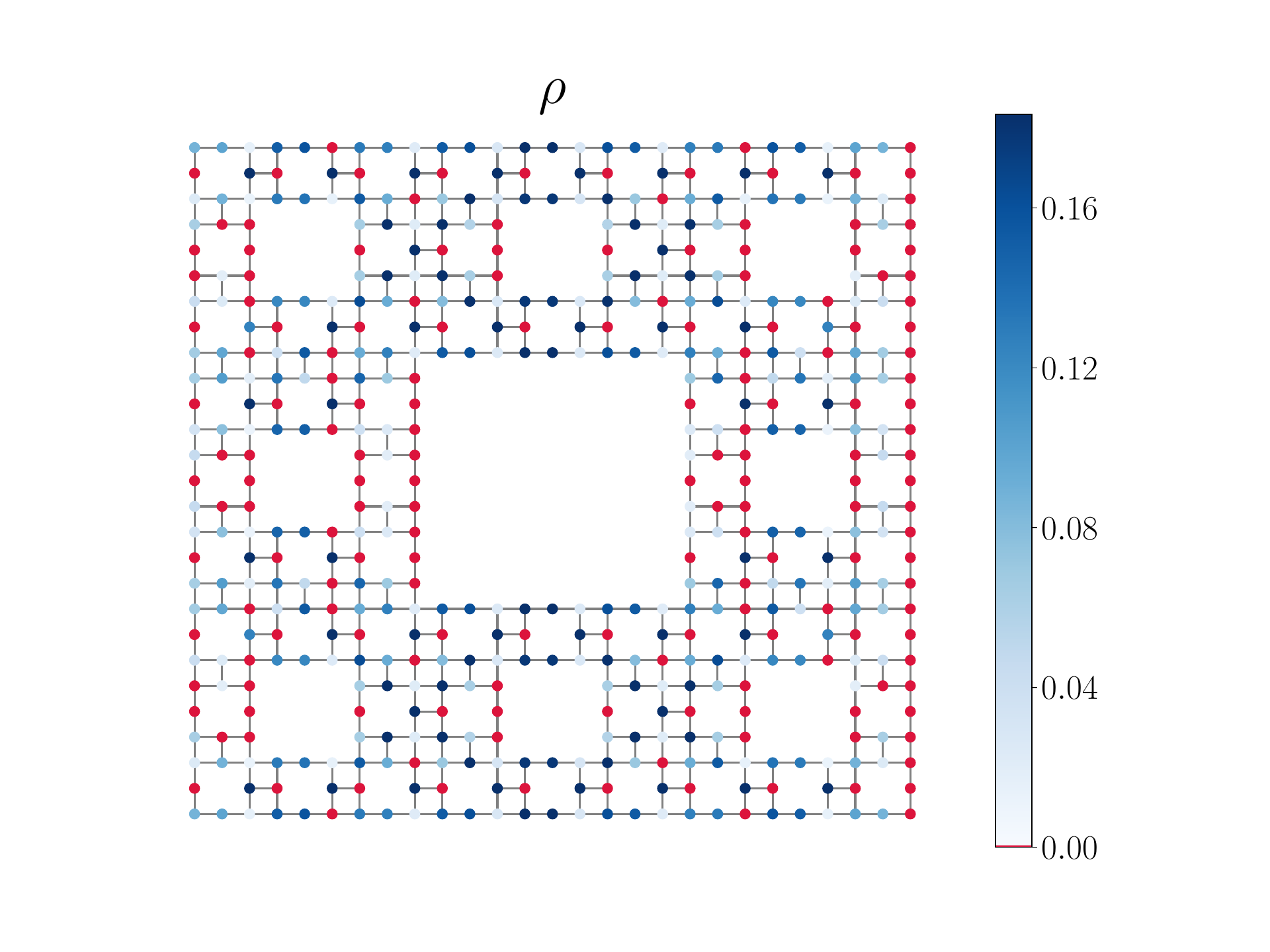}
    \caption{\label{fig:rho_square} The superfluid density $\rho=D_s/\pi$ at $U=1.0$, chemical potential $\mu=0.0$, temperature $T=0.005$ for the top figures, and $T=0.025$ (around $T_c$ of the undeformed lattice) for the bottom ones. Two structures are shown -- square sample with the disorder ($30\%$ of deleted sites) and the Sierpi\'nski carpet of depth $3$. The sites with $D_s/\pi<0.01$ are indicated by red color.}
\end{figure*}

While there seems to be no enhancement of superconductivity on the Sierpi\'nski carpet as compared with the square lattice case, for the sake of completeness we plot spatial profiles of the Cooper pair condensate and the superfluid stiffness on the carpet and on the randomly disordered lattice with approximately the same number of removed sites. Those are shown in Figs. \ref{fig:delta_square} and \ref{fig:rho_square}.

\section*{Supplementary Note 4: critical temperature dependence on the density of states}
Finally, we provide relations between the critical temperature of superconducting transition $T_c$ and the corresponding density of states at the Fermi level $g(E_F)$. Here, we stick to the case of $U=1$ considered in the main text. For that, at each value of $\mu$, $T_c$ was estimated as the temperature at which the average value of superconducting condensate drops below $\Delta < 0.001$, and the corresponding density of states at the Fermi level has been computed.

\begin{figure*}[h!]
    \centering
    \includegraphics[width=0.45\textwidth]{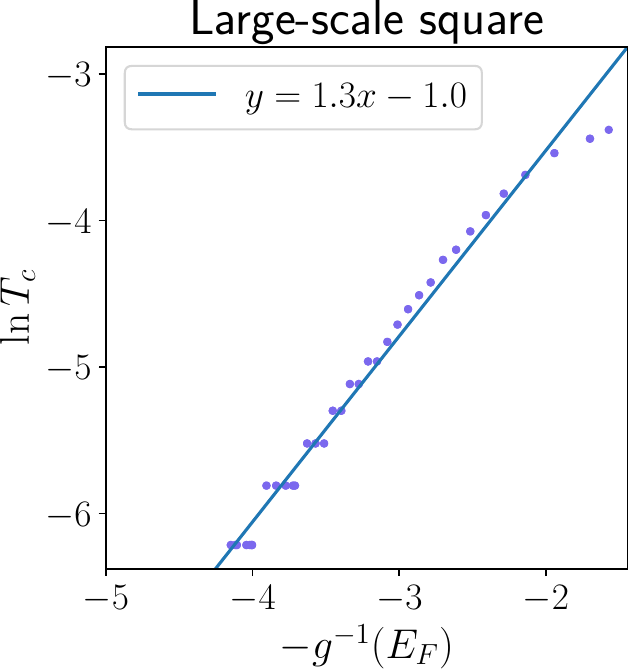}
    \includegraphics[width=0.45\textwidth]{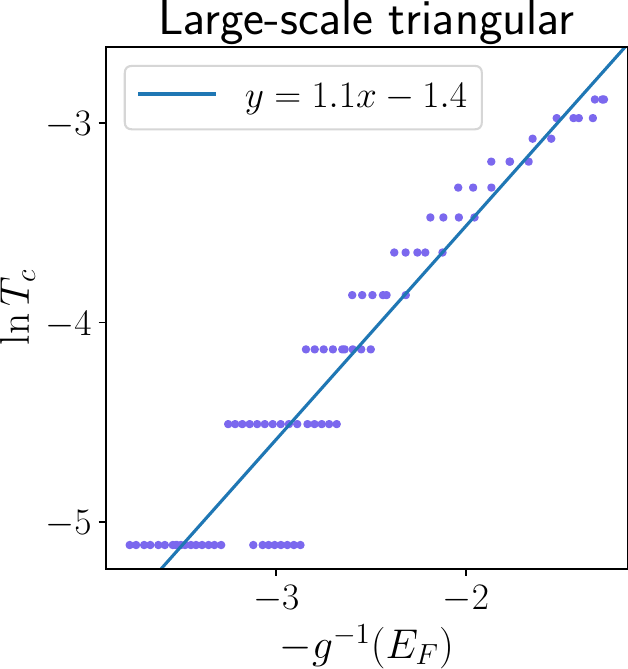}
    \caption{\label{fig:Tc_gE_th} Dependence of the logarithm of the critical temperature on the (inverse) density of states at the Fermil level for the square and triangular lattices in the thermodynamic limit.}
\end{figure*}

\begin{figure*}[h!]
    \centering
    \includegraphics[width=0.45\textwidth]{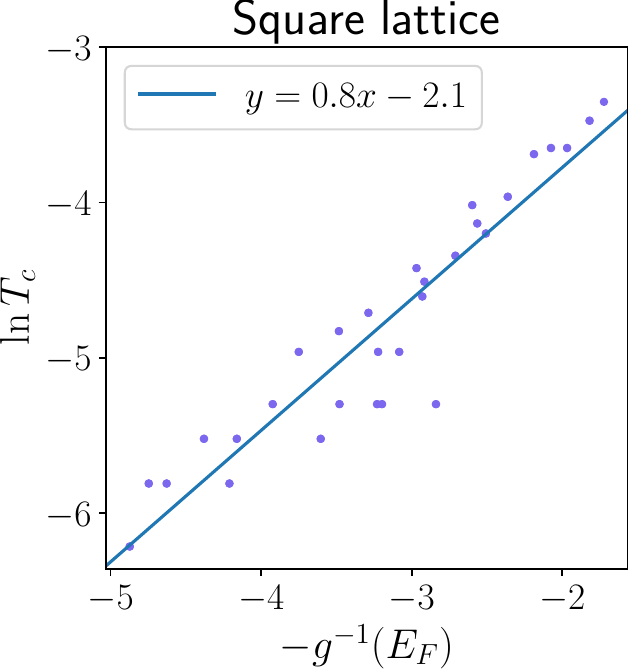}
    \includegraphics[width=0.45\textwidth]{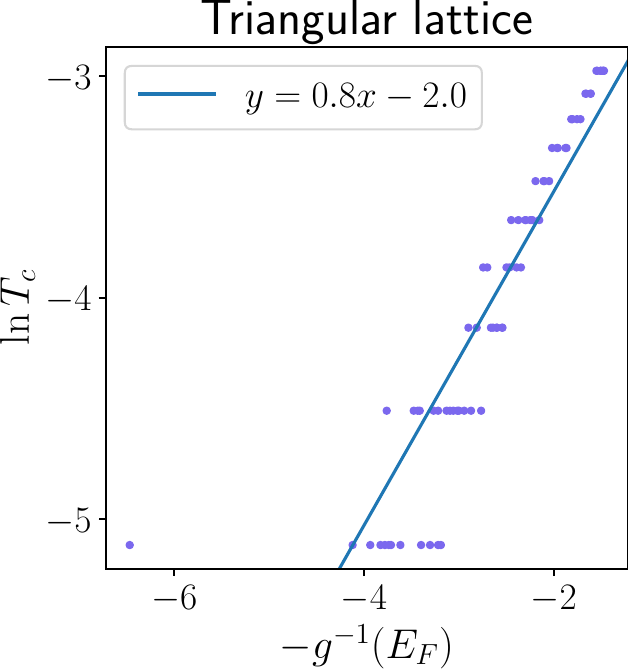}
    \includegraphics[width=0.45\textwidth]{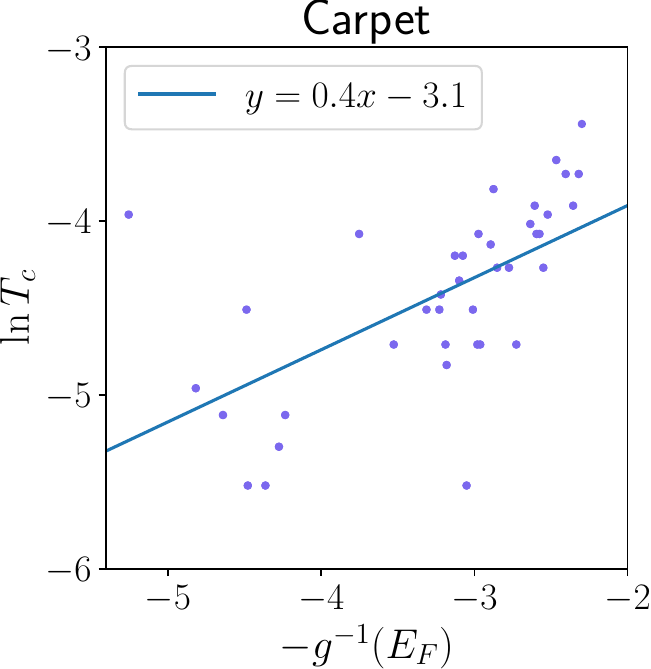}
    \includegraphics[width=0.45\textwidth]{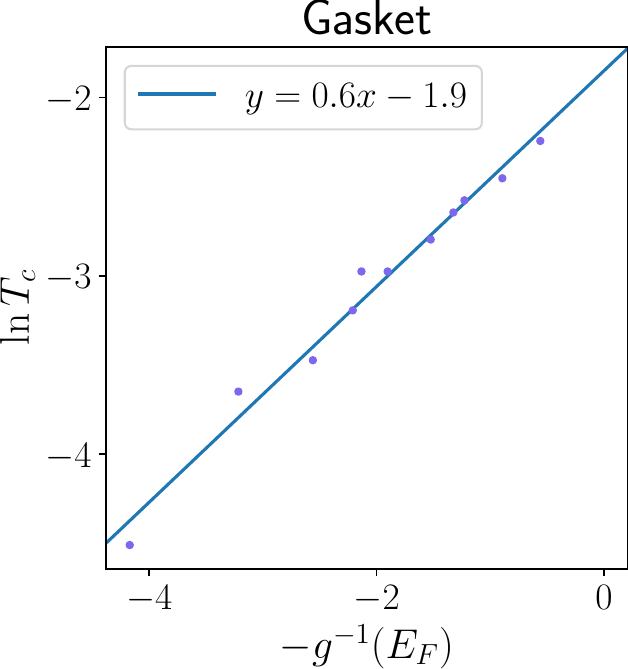}
    \caption{\label{fig:Tc_gE} Dependence of the logarithm of the critical temperature on the (inverse) density of states at the Fermil level for finite-size square and triangular lattices, and for the corresponding Sierpinski carpet and gasket structures.}
\end{figure*}

These dependencies are shown in Figs. \ref{fig:Tc_gE_th} and \ref{fig:Tc_gE} in log-linear scale in the thermodynamic limit and in the finite-size case correspondingly. For the square, triangular, and the Sierpinski gasket lattices, one can see rather clean linear relations, corresponding to the BCS theory prediction $T_c \simeq \theta e^{-\alpha / U g(E_F)}$. In the BCS theory $\alpha=1$, while in the considered cases, it varies between $0.8-1.3$ for the regular lattice, and can drop to $0.6$ for the Sierpinski gasket, which can be due its unconventional geometric properties. For the Sierpinski carpet structure, the results are much more noisy and less conclusive, which comes from the fact that the phase diagram (Fig. 1 in the main text, ``Sierpinski carpet'' panel) is rather ragged, and, given finite numerical precision, it is difficult to accurately determine $T_c$ for each value of $\mu$.

Overall, the BCS relation between $T_c$ and $g(E_F)$ is satisfied, and the elevation of $T_c$ for the Sierpinski gasket can be attributed to the increase of density of states at the Fermi energy around the optimal doping, as well as to the decrease in $\alpha$ parameter. 

